\documentclass[journal,twocolumn]{IEEEtran}
\usepackage{epsfig,rotating,setspace,latexsym,amsmath,epsf,amssymb,amsfonts,bm,theorem,cite,algorithm,algorithmic,graphicx,epsf,authblk,epstopdf,color,accents,amssymb,subfigure}

\IEEEoverridecommandlockouts
\allowdisplaybreaks
\DeclareMathOperator*{\argmax}{arg\,max}
\DeclareMathOperator*{\argmin}{arg\,min}

\newtheorem{theorem}{Theorem}
\newtheorem{lemma}{Lemma}
\newenvironment{Proof}[1]{\medskip\par\noindent{\bf Proof:\,}\,#1}{{\mbox{\,$\blacksquare$}\par}}

\begin{document}

\title{The Freshness Game: Timely Communications \\ in the Presence of an Adversary}

\author{Subhankar Banerjee,~\IEEEmembership{Graduate Student Member,~IEEE,}  \qquad Sennur Ulukus,~\IEEEmembership{Fellow,~IEEE}
\thanks{S. Banerjee and S. Ulukus are with the Department of Electrical and Computer Engineering, University of Maryland, College Park, MD 20742 (e-mails: sbanerje@umd.edu and ulukus@umd.edu).}}


\maketitle

\begin{abstract}
We consider a communication system where a base station (BS) transmits update packets to $N$ users, one user at a time, over a wireless channel. We investigate the age of this status updating system with an adversary that jams the update packets in the downlink. We consider two system models: with diversity and without diversity. In the model without diversity, in each time slot, the BS schedules a user from $N$ users according to a user scheduling algorithm. The constrained adversary blocks at most a given fraction, $\alpha$, of the time slots over a horizon of $T$ slots, i.e., it can block at most $\alpha T$ slots of its choosing out of the total $T$ time slots. We show that if the BS schedules the users with a stationary randomized policy, then the optimal choice for the adversary is to block the user which has the lowest probability of getting scheduled by the BS, at the middle of the time horizon, consecutively for $\alpha T$ time slots. The interesting \emph{consecutive property} of the blocked time slots is due to the cumulative nature of the age metric. In the model with diversity, in each time slot, the BS schedules a user from $N$ users and chooses a sub-carrier from $N_{sub}$ sub-carriers to transmit update packets to the scheduled user according to a user scheduling algorithm and a sub-carrier choosing algorithm, respectively. The adversary blocks $\alpha T$ time slots of its choosing out of $T$ time slots at the sub-carriers of its choosing. We show that for large $T$, the uniform user scheduling algorithm together with the uniform sub-carrier choosing algorithm is $\frac{2 N_{sub}}{N_{sub}-1}$ optimal. Next, we investigate the game theoretic equilibrium points of this status updating system. For the model without diversity, we show that a Nash equilibrium does not exist, however, a Stackelberg equilibrium exists when the scheduling algorithm of the BS acts as the leader and the adversary acts as the follower. For the model with diversity, we show that a Nash equilibrium exists and identify the Nash equilibrium. Finally, we extend the model without diversity to the case where the BS can serve multiple users and the adversary can jam multiple users, at a time. 
\end{abstract}
 
\begin{IEEEkeywords}
Age of information, information freshness, jamming adversary, game theory.
\end{IEEEkeywords}

\section{Introduction}
Due to the advances in modern applications such as virtual reality, augmented reality, robotics, mission critical control and various other 5G technologies, the \emph{quality of user experience} has become important, perhaps even more important than \emph{quality of service}, which is typically measured in terms of delay, throughput and bit error. A recently introduced quality of user experience metric is \emph{freshness of information} which is measured in terms of the \emph{age of information} \cite{kaul2012real}. In a slotted system, the age of information of a user is $t-\tau$, where $t$ is the current time slot and $\tau$ is the last time slot when this user has received an update packet. Maintaining information freshness at a user is different than maintaining low delay or high throughput at the user; fresh information delivery is achieved by delivering sufficiently frequent packets at sufficiently low delays, which requires operating the system at a novel interior throughput-delay point; see recent surveys \cite{kosta2017age, SunSurvey, YatesSurvey}. 

A vast amount of literature is available now on the analysis and optimization of the age of information metric ranging from various queueing regimes to energy harvesting systems, wireless networks, remote estimation, gossip networks, caching systems, source coding problems, and so on, see e.g., \cite{Najm17, Soysal19, Ayan19, Yates20_moments, Farazi18, Wu18, Baknina18, Leng19, Arafa19TwoHop, gu2020twohop, Arafa20, wireless-ephremides, kadota18, Buyukates19_multihop, hsu18, Buyukates19_hier, Elmagid20, Ceran18, Buyukates20_stragglers, zou19,  yang20, ozfatura20, buyukates-fl, liu18, wang19_counting, bastopcu20_google, sun17_remote, chakravorty20, Bastopcu19_distortion, vaze21jsac, Bastopcu20_soft_updates, Yates_Soljanin_source_coding, mayekar20, Bastopcu20_selective, Yates17sqrt, bastopcu2020LimitedCache, Eryilmaz21, Kaswan-isit2021, Yates21gossip, baturalp21comm_struc, bastopcu21gossip, kaswan2022age, kaswantimestomp, kaswanslice, mitraasuman, kam20}. Most of the existing literature considers communication systems without any adversaries. The recent works that are most closely related to our work here are \cite{banerjee2020fundamental, bhattacharjee2020competitive, garnaev2019maintaining, nguyen2017impact, xiao2018dynamic, banerjee-infocom2022, banerjee2022game}. Specifically, \cite{banerjee2020fundamental, bhattacharjee2020competitive} use an adversary in their communication system to model non-stationarity of a wireless communication channel. The adversary blocks a communication channel by completely eliminating the transmitted update packet, increasing the age. Similar to \cite{banerjee2020fundamental, bhattacharjee2020competitive}, we consider an adversary which blocks a communication link by completely eliminating the update packets. However, different than \cite{banerjee2020fundamental, bhattacharjee2020competitive}, we consider an energy constrained adversary; in addition, while \cite{banerjee2020fundamental, bhattacharjee2020competitive} consider competitive ratio, we consider the absolute age performance. \cite{garnaev2019maintaining, nguyen2017impact} consider an adversary that acts as an interferer which decreases the signal-to-noise-ratio of the communication link, decreasing the data rate, ultimately increasing the age. \cite{xiao2018dynamic} considers an adversary which blocks a communication channel for a time duration in continuous time, which results in higher age for the system, by disabling the transmissions for some time after the reception of the last update packet. Unlike \cite{garnaev2019maintaining, nguyen2017impact, xiao2018dynamic}, our considered adversary increases the age by directly jamming (blocking) the channel in time slots of its choosing in a time slotted system.

\begin{figure}[t]
	\centerline{\includegraphics[width = 0.6 \columnwidth]{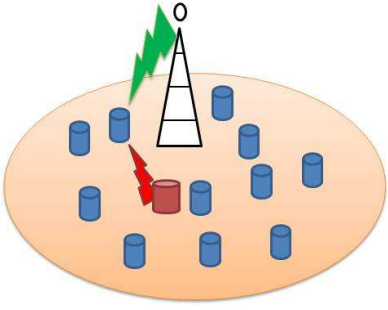}}
	\caption{A base station serves users, and an adversary blocks communication.}
	\label{fig1}
	\vspace*{-0.4cm}
\end{figure}

In this paper, the adversary can block any $\alpha T$ time slots over a time horizon of $T$ slots where $\alpha < 1$. First, we investigate the system where, at every time slot, the base station (BS) schedules a user from $N$ users and the adversary blocks any one of the $N$ users for any $\alpha T$ time slots, see Fig.~\ref{fig1}. In this case, we do not have any diversity present in the system. For this case, we show that if the BS employs any stationary randomized policy then the optimal adversarial action is to block the user which has the lowest probability of getting scheduled by the BS, in the middle of the time horizon for consecutive $\alpha T$ time slots. We also show that the Nash equilibrium does not exist for this system. However, a Stackelberg equilibrium exists, and we find the Stackelberg equilibrium point for this system. Next, we investigate the system with diversity by introducing diversity in frequency domain via sub-carriers \cite{bahai2004multi}. In this system, at every time slot, the BS schedules a user from $N$ users and chooses a sub-carrier among $N_{sub}$ sub-carriers. The adversary blocks any one of the $N_{sub}$ sub-carriers for any $\alpha T$ time slots. For this case, we show that the Nash equilibrium exists, and we find the Nash equilibrium point. Finally, we extend the model without diversity to the case where the BS and the adversary can serve multiple users and jam multiple users, respectively, at a time. 

\section{System Model and Problem Formulation}
We consider two different settings for a wireless communication network. In the first setting, there is no diversity, and in the second setting, we incorporate diversity in the frequency domain by introducing multiple sub-carriers. In both settings, a BS aims to minimize the age of information of the network by regular transmission of the update packets to the users present in the system. In both settings, an adversary present in the system aims to prevent the transmission of the update packets by jamming them to maximize the age of the system. 

At time slot $t$, the BS has full knowledge about the adversarial actions till time slots $t'$, where $0\leq t'<t$, however the BS has no knowledge about any adversarial actions for time slots $t'$, where $t \leq t' \leq T$. The adversary has full knowledge about the algorithm employed by the BS, however if the BS uses a randomized algorithm, at time slot $t$, the adversary does not know the specific action of the BS for time slot $t'$. At time slot $t$ the adversary can optimize its action for time slot $t'$ based on the available information till time slot $t$, where $t\leq t'\leq T$. 

We provide an example to illustrate these assumptions. Let us assume that at every time slot, the BS schedules users with a probability mass function (pmf) $\bm{p}=[p_1 ~ p_2 ~ \cdots ~ p_N]$, i.e., the BS schedules user $i$ with probability $p_{i}$. At time slot $t$, the adversary has no knowledge about which user is going to get scheduled at time slot $t'$, where $t\leq t'\leq T$, however, the adversary knows that the BS is scheduling the users following pmf $\bm{p}$. We assume that at each time slot, the BS has a fresh update packet to transmit for every user present in the system. As we are interested in the freshness, at each time slot, all the undelivered update packets get dropped at the BS without any cost, which is a valid assumption in the literature \cite{kadota18,banerjee2020fundamental}. 

\subsection{Communication Network Model Without Diversity}\label{subsec:A}
\begin{figure}[t]
    \centerline{\includegraphics[width = 0.8 \columnwidth]{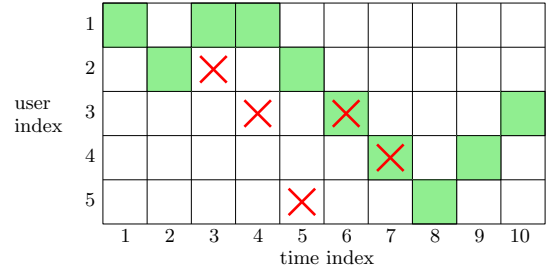}}
    \caption{The base station serves users in green shaded time slots, and the adversary blocks communication in red crossed time slots.}
    \label{fig2}
    \vspace*{-0.4cm}
\end{figure}

We consider a communication system, where at each time slot a BS schedules a user from $N$ users with a scheduling policy $\pi_{u}$. The BS transmits an update packet to the scheduled user at each time slot. The adversary present in the system can block a communication channel between a user and the BS. By blocking a communication channel, the adversary completely eliminates the update packet. The adversary can block only one communication channel at a given time slot, and in total it can block $\alpha T$ time slots over the time horizon of $T$ ($T>0$) time slots, where $0<\alpha<1$. Let $\sigma_{i}(t)$ denote the action of the adversary for the communication channel between user $i$ and the BS at time $t$. Here $\sigma_{i}(t) = 0$ means that the adversary blocks user $i$ at time $t$, and $\sigma_{i}(t)=1$ means that the adversary does not block user $i$ at time $t$. Thus, the action of the adversary against user $i$ is a sequence of ones and zeros. We denote this sequence with $\sigma_{i}$, and we call this sequence a \emph{blocking sequence} for user $i$. We use $\sigma$ to denote the blocking matrix whose $i$th row is the blocking sequence $\sigma_{i}$, and thus, $(i,t)$th entry $\sigma_i(t)$ is the state of the communication channel between user $i$ and the BS at time $t$. Thus, a feasible $\sigma$ should satisfy $\sum_{i=1}^{N}\sum_{t=1}^{T} (1\!-\!\sigma_i(t)) \leq \alpha T$. We create a set $\Sigma$ of feasible $\sigma$. A pictorial representation of the adversarial action is given in  Fig.~\ref{fig2}.

Formally, age for a user scheduling algorithm $\pi_{u}$ (run by the BS) and a blocking matrix $\sigma$ (run by the adversary) is
\begin{align} \label{objective}
\Delta^{\pi_{u},\sigma} = \frac{1}{T} \sum_{t=1}^{T} \frac{1}{N} \left(\sum_{i=1}^{N} \mathbb{E}\left[a_{i}(t)\right]\right)
\end{align}
where $a_{i}(t)$ is the age of the $i$th user at time $t$. Note that if the BS successfully transmits an update packet to the $i$th user at time $t$ then $a_{i}(t+1)$ becomes $1$, otherwise $a_{i}(t+1)$ becomes $a_{i}(t)+1$.  We denote $\Delta_{i}(t)=E[a_{i}(t)]$, the average age of the $i$th user at time $t$. We also denote the overall age of user $i$ as
\begin{align}\label{eq:new2}
\Delta_i=\frac{1}{T} \sum_{t=1}^T \Delta_i(t)
\end{align}
 
The adversary aims to increase the age and the BS aims to decrease the age of the system. We study the interaction between the adversary and the algorithm via the problem,
\begin{align}\label{eq:prob_formu}
      \Delta^{*} = \sup_{\sigma} \inf_{\pi_{u}} \quad & \Delta^{\pi_{u},\sigma} \nonumber \\  
       \textrm{s.t.}  \quad & \sum_{i=1}^{N} \sum_{j=1}^{T} 
       (1 - \sigma_i(j)) \leq \alpha T\nonumber\\ 
      & \sum_{i=1}^{N} (1-\sigma_i(j)) \leq 1, \quad j=1,\ldots,T
\end{align}
The first constraint in (\ref{eq:prob_formu}) is due to the energy constraint of the adversary, and the second constraint is due to the fact that at any given time slot the adversary can block at most 1 user. For simplicity of notation, we drop $\pi_{u}$ and $\sigma$ from the superscript of $\Delta$ in the rest of the paper, unless we specify otherwise. 
 
Next, we investigate the game theoretic equilibrium points for this system model when the action space of the BS is limited to stationary randomized policies, i.e., the BS schedules a user from $N$ users following a stationary distribution. Let $p_{i}$ be the probability with which the BS schedules user $i$. Thus, the average time needed for the user $i$ to get scheduled is $\frac{1}{p_{i}}$. Hence, we assume that the time horizon $T\gg \frac{1}{p_{i}}$ for all $i$. In addition, we assume that $T$ is large enough such that for a given $\alpha$, $T(1-\alpha)$ is also large.

Let $\bm{p}=[p_1 ~ p_2 ~ \cdots ~ p_N]$ be the pmf with which the BS schedules users at every time slot. For the stationary regime of scheduling polices, we re-denote (\ref{objective}), for a given scheduling pmf $\bm{p}$ and a blocking matrix $\sigma$, as  
\begin{align} \label{objective1}
\Delta^{\bm{p},\sigma} =  \frac{1}{T} \sum_{t=1}^{T} \frac{1}{N} \left(\sum_{i=1}^{N} \mathbb{E}\left[a_{i}(t)\right]\right)
\end{align}
For a given scheduling pmf $\bm{p}$, the adversary aims to maximize $\Delta^{\bm{p},\sigma}$. We define a set $B(\bm{p})$, which consists of the solutions to the following optimization problem,
\begin{align}
    B(\bm{p}) = \argmax_{\sigma\in{\Sigma}} \Delta^{\bm{p},\sigma}
\end{align}
Similarly, for a given adversarial action $\sigma$, the BS aims to minimize $\Delta^{\bm{p},\sigma}$. We define a set $B(\sigma)$, which consists of the solutions to the following optimization problem,
\begin{align}
    B(\sigma) = \argmin_{\bm{p}\in{\mathcal{F}}} \Delta^{\bm{p},\sigma}
\end{align}
where $\mathcal{F}$ is the set of feasible pmfs. Here, $B(\bm{p})$ and $B(\sigma)$ denote the \emph{best responses} to actions $\bm{p}$ and $\sigma$, respectively.
 
A scheduling algorithm and an adversarial action, $(\bar{\bm{p}},\bar{\sigma})$ form a Nash equilibrium point \cite{han2019game} if and only if $\Delta^{\bar{\bm{p}}, \bar{\sigma}} \leq \Delta^{\hat{\bm{p}},\bar{\sigma}}$ and $\Delta^{\bar{\bm{p}},\bar{\sigma}}\geq\Delta^{\bar{\bm{p}},\hat{\sigma}}$, for all $\hat{\bm{p}}\in{\mathcal{F}}$ and for all $\hat{\sigma}\in \Sigma$. In other words, $(\bar{\bm{p}},\bar{\sigma})$ is a Nash equilibrium point if and only if $\bar{\bm{p}}\in B(\bar{\sigma})$ and $\bar{\sigma}\in B(\bar{\bm{p}})$. 
 
We also provide a Stackelberg equilibrium point for this system model when the scheduling algorithm of the BS acts as a leader. Formally, $(\bar{\bm{p}},\bar{\sigma})$ is a Stackelberg equilibrium point \cite{han2019game} if and only if $\bar{\sigma}\in B({\bar{\bm{p}}})$ and $\bar{\bm{p}} \in \argmin_{\bm{p}\in\mathcal{F}} \Delta^{\bm{p},{\sigma}|{\sigma}\in{B(\bm{p})}}$.

In Section~\ref{sec:5}, we consider a general setting for this model, where at each time slot, the BS schedules $k$ users and the adversary can block $k_{a}$ users with a constraint that it can block maximum $\alpha T$ times. Thus, the constraints for the adversary are modified as follows,
\begin{align}
    &\sum_{i=1}^{N}\sum_{j=1}^{T} 
       (1 - \sigma_i(j)) \leq \alpha T\nonumber\\ 
      & \sum_{i=1}^{N} (1-\sigma_i(j)) \leq k_{a}, \quad j=1,\ldots,T
\end{align}

\subsection{Communication Network Model with Diversity}\label{sub:2}
We consider a communication system, where at every time slot, the BS schedules a user from $N$ users with a user scheduling algorithm $\pi_{u}$, and chooses a sub-carrier from available $N_{sub}$ sub-carriers $(N_{sub}>1)$ with a sub-carrier choosing algorithm $\pi_{s}$ to transmit update packets to the scheduled user. At a given time slot, the adversary present in the system can block a sub-carrier out of $N_{sub}$ sub-carriers, and in total it can block $\alpha T$ sub-carriers over the time horizon $T$, where $0<\alpha < 1$. Similar to the previous setting, blocking a sub-carrier implies complete elimination of the update packet.

\begin{figure}[t]
    \centerline{\includegraphics[width = 0.8\columnwidth]{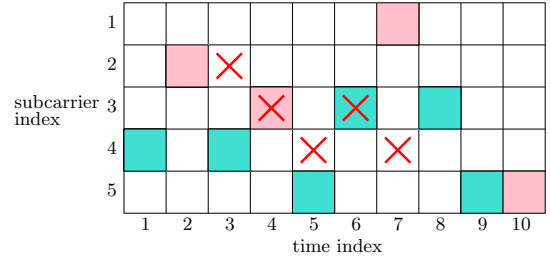}}
    \caption{A pictorial illustration of the modified system model: Pink and turquoise colors represent two users. In each time slot, the BS chooses one user and one sub-carrier to send an update. For example, at time slot $5$, the BS chooses turquoise user, and the $5$th sub-carrier to send an update; and the adversary chooses to block the $4$th sub-carrier.}
    \label{fig8}
    \vspace*{-0.4cm}
\end{figure}

We denote the action of the adversary for the $j$th sub-carrier at time slot $t$ as $\sigma_{j}(t)$. Here, $\sigma_{j}(t)=0$ means that the adversary blocks sub-carrier $j$ at time $t$, and $\sigma_{j}(t)=1$ means that the adversary does not block sub-carrier $j$ at time $t$. Thus, the action of the adversary to sub-carrier $j$ is a sequence of ones and zeros, and we denote this sequence as $\sigma_{j}$, and we call this sequence as the blocking sequence for sub-carrier $j$. We use $\sigma$ to denote the blocking matrix, whose $j$th row is the blocking sequence $\sigma_{j}$, for the $j$th sub-carrier, thus, the $(j,t)$th entry of $\sigma$, $\sigma_j(t)$ denotes the state of the communication channel on sub-carrier $j$ at time $t$. A pictorial representation of the adversarial action is given in Fig.~\ref{fig8}. Thus, a feasible $\sigma$ should satisfy $\sum_{j=1}^{N_{sub}}\sum_{t=1}^{T} (1-\sigma_j(t)) \leq \alpha T$. We create a set $\bar{\Sigma}$ of feasible $\sigma$. In the previous setting, we have a dedicated communication channel between a user and the BS, however, in the current setting, there is a pool of communication channels, namely, $N_{sub}$ communication channels, and at every time slot, the BS chooses one of these $N_{sub}$ communication channels with a sub-carrier choosing algorithm $\pi_{s}$. 

We define the average expected age of the system for a user scheduling algorithm $\pi_{u}$, sub-carrier choosing algorithm $\pi_{s}$ (both are run by the BS) and a blocking matrix $\sigma$ (run by the adversary) as
\begin{align}\label{eq:new7}
\Delta^{\pi_{u}, \pi_{s}, \sigma} = \frac{1}{T} \sum_{t=1}^{T} \left(\frac{1}{N} \sum_{i=1}^{N} \mathbb{E}[a_{i}(t)] \right)
\end{align}

We study the interaction between the BS and the adversary as the following optimization problem,
\begin{align}\label{eq:prob_formu2}
      \Delta^{*} = \sup_{\sigma} \inf_{\pi_{u}, \pi_{s}} \quad & \Delta^{\pi_{u}, \pi_{s}, \sigma} \nonumber \\  
       \textrm{s.t.}  \quad & \sum_{k=1}^{N_{sub}} \sum_{j=1}^{T} (1 -\sigma_k(j)) \leq \alpha T\nonumber\\ 
      & \sum_{k=1}^{N_{sub}} (1 - \sigma_k(j)) =1, \quad j=1,\ldots,T
\end{align}

Next, we find the Nash equilibrium point for this system model, when the action space of the BS is limited to stationary randomized policies, i.e., the BS schedules a user from $N$ users following a stationary distribution and chooses a sub-carrier from $N_{sub}$ sub-carriers following a stationary distribution. Let $\bm{p}$ be the pmf with which the BS schedules a user at every time slot, and let $\bm{q}$ be the pmf with which the BS chooses a sub-carrier at every time slot to transmit the update packet to the scheduled user, and let the adversary employ a blocking matrix $\sigma$. Then, according to (\ref{eq:new7}) the average expected age of the system becomes
\begin{align}
\Delta^{\bm{p}, \bm{q}, \sigma} = \frac{1}{T} \sum_{t=1}^{T} \left(\frac{1}{N} \sum_{i=1}^{N} \mathbb{E}[a_{i}(t)] \right)
\end{align}

For a given user scheduling pmf $\bm{p}$ and a sub-carrier selection pmf $\bm{q}$, the adversary aims to maximize $\Delta^{\bm{p},\bm{q},\sigma}$. We define a set $B(\bm{p},\bm{q})$, which consists of the solutions to the following optimization problem
\begin{align}
    B(\bm{p},\bm{q}) = \argmax_{\sigma\in{\bar{\Sigma}}} \Delta^{\bm{p},\bm{q},\sigma}
\end{align}
Similarly, for a given adversarial action $\sigma$, the BS aims to minimize $\Delta^{\bm{p}, \bm{q}, \sigma}$. We define a set $B(\sigma)$, which consists of the solutions to the following optimization problem
\begin{align}
    B(\sigma) = \argmin_{\bm{p}\in{\mathcal{F}}, \bm{q}\in{\mathcal{F}'}} \Delta^{\bm{p}, \bm{q},\sigma}
\end{align}
where $\mathcal{F}$ is the set of feasible probability distributions of choosing a user from $N$ users, and $\mathcal{F}'$ is the set of probability distributions of choosing a sub-carrier from $N_{sub}$ sub-carriers. Then, $(\bar{\bm{p}},\bar{\bm{q}},\bar{\sigma})$ is a Nash equilibrium point if and only if $(\bar{\bm{p}},\bar{\bm{q}})\in{B(\bar{\sigma})}$ and $\bar{\sigma}\in{B(\bar{\bm{p}},\bar{\bm{q}})}$. We show that Nash equilibrium for this setting exists, and find the Nash equilibrium. For simplicity of notation, we drop any superscripts of $\Delta$ in the rest of the paper, unless we specify otherwise.

\section{Analysis of Age in Model Without Diversity}
We begin this section with two simple lemmas. Lemma~\ref{lemma:1} below states that any deterministic user scheduling algorithm, which chooses the order in which users are served by the BS deterministically, achieves an age that is lower bounded by a function that increases linearly with $T$ that does not depend on $N$. Lemma~\ref{lemma:2} below, on the other hand, states that if the algorithm chooses the user to be served by the BS uniformly at random, even though the average age is again lower bounded by a term that increases linearly with $T$, the coefficient of the linear increase is inversely proportional with $N$. That is, if the number of users $N$ is large, then the age can decrease to a satisfactory level, despite the presence of an adversary. 

\begin{lemma}\label{lemma:1}
 For any deterministic user scheduling algorithm, the age is lower bounded by $\frac{T \alpha^2}{2}$.
\end{lemma}

\begin{Proof}
Since the algorithm is deterministic, the adversary knows which user is being served in each time slot. Since we are finding a lower bound, we consider that the adversary blocks consecutive $\alpha T$ time slots, blocking the user being served in each time slot. During this blocking time interval, the ages of all users increase, because no one is served. Since we are finding a lower bound, assume that the ages of all users start from 1, i.e., the ages of all users increase from 1 to $\alpha T$ in this interval, and also, assume that the ages in the remaining $(1-\alpha)T$ slots are all 1. Thus, age user $i$ is lower bounded as,
\begin{align}
    \Delta_{i} \geq \frac{1}{T} \left( \sum_{t=1}^{\alpha T} t + \sum_{t=1}^{(1-\alpha) T} 1 \right) \geq \frac{T\alpha^{2}}{2} + (1-\alpha)
\end{align}
Thus, the overall age is lower bounded as
\begin{align}
    \Delta = \frac{1}{N} \sum_{i=1}^{N}\Delta_{i} \geq \frac{T\alpha^{2}}{2} + (1-\alpha) \geq \frac{T\alpha^{2}}{2}
\end{align}
which is linear in $T$.
\end{Proof}
 
\begin{lemma} \label{lemma:2}
For any randomized user scheduling algorithm, the age is lower bounded by $\frac{T\alpha^{2}}{2 N }$.    
\end{lemma}

\begin{Proof}
Since we are finding a lower bound, we consider that the adversary blocks consecutive $\alpha T$ time slots of a fixed user. Thus, for these $\alpha T$ time slots, one of the users' age increases from 1 to $\alpha T$ assuming that it starts at 1, also that the ages of all users anywhere else is 1, both of which are due to the fact that we are finding a lower bound. The age of the blocked user is lower bounded as 
\begin{align}
    \Delta_{i} \geq \frac{1}{T} \left( \sum_{t=1}^{\alpha T} t + \sum_{t=1}^{(1-\alpha) T} 1 \right) \geq \frac{T\alpha^{2}}{2} + (1-\alpha)
\end{align}
and the ages of all other users are lower bounded as
\begin{align}
    \Delta_{k} \geq \frac{1}{T} \left( \sum_{t=1}^{T} 1 \right) \geq 1
\end{align}
Thus, the overall age is lower bounded as
\begin{align}
\Delta = \frac{1}{N} \sum_{i=1}^{N}\Delta_{i} \geq \frac{1}{N} \left(\frac{T\alpha^{2}}{2} + (1-\alpha) +(N-1) \right) \geq \frac{T\alpha^{2}} {2 N}    
\end{align}
which is linear in $T$, but is inverse linear in $N$.
\end{Proof}

The lower bounds in Lemmas~\ref{lemma:1}~and~\ref{lemma:2} hint at the possibility that we may be able to achieve a lower average age for the system by using a randomized scheme, in the presence of an adversary. In the rest of this paper, we will carefully analyze the age performance of a randomized algorithm in the presence of an adversary. First, we will show that, when the algorithm is random, an optimal adversary blocks $\alpha T$ consecutive slots of a randomly chosen user. We find this \emph{consecutive} nature of the optimum blocking strategy of the adversary noteworthy, which is inherent to the age metric. For instance, if the metric was throughput, an optimal adversary would block any randomly chosen $\alpha T$ time slots; whether the blocked time slots are consecutive or not does not have any bearing on the throughput metric, while it is crucial for the age metric.

To analyze the performance of randomized algorithms in the presence of an adversary, we first note the following facts: If the algorithm does not choose the $i$th user at time $t$, then irrespective of $\sigma_{i}(t)$, $a_{i}(t)$ increases by $1$, i.e., $a_i(t+1)=a_i(t)+1$. The probability of this event is $\frac{N-1}{N}$. If the algorithm chooses the $i$th user at time $t$, then $a_{i}(t+1)$ depends on $\sigma_{i}(t)$. If $\sigma_{i}(t)=0$, i.e., the adversary blocks user $i$, then $a_{i}(t+1)=a_{i}(t)+1$. If $\sigma_{i}(t)=1$, i.e., the adversary does not block user $i$, then $a_{i}(t+1)=1$. Thus, if the algorithm chooses the $i$th user at time $t$, we can write $a_{i}(t+1)$ in a compact form as $a_{i}(t+1)=a_{i}(t)(1-{\sigma}_{i}(t))+1$. The probability of the algorithm choosing user $i$ is $\frac{1}{N}$. Thus, the expected age for user $i$ at time $t+1$ conditioned on the age at time $t$ is
\begin{align}\label{eq:basic}
 \mathbb{E}\left[a_{i}(t+1)|a_{i}(t)\right] = & \frac{N-1}{N} \left(a_{i}(t)+1\right) \nonumber\\ 
 & + \frac{1}{N} \left(a_{i}(t)(1-{\sigma}_i(t))+1\right)   
\end{align}
which simplifies to
\begin{align}\label{eq:7}
 \mathbb{E}[a_{i}(t+1)|a_{i}(t)] = a_{i}(t)\left(1-\frac{{\sigma}_{i}(t)}{N}\right)+ 1
\end{align}
Note that, 
\begin{align}\label{eq:8}
    \mathbb{E}\left[\mathbb{E}\left[a_{i}(t+1)|a_{i}(t)\right]\right] = \Delta_{i}(t+1)
\end{align}
Then, (\ref{eq:7}) and (\ref{eq:8}) give
\begin{align}\label{eq:9}
 \Delta_i(t+1) = \Delta_i(t)\left(1-\frac{{\sigma}_{i}(t)}{N}\right)+ 1
\end{align}
showing how age of user $i$, $\Delta_i(t)$, evolves from time $t$ to $t+1$ as a function of $\sigma_i(t)$.  
Writing one more recursion, we obtain
\begin{align}\label{eq:10}
 \Delta_i(t+1) = & \Delta_i(t-1)\left(1-\frac{{\sigma}_{i}(t-1)}{N}\right)\left(1-\frac{{\sigma}_{i}(t)}{N}\right) \nonumber\\
 &+ \left(1-\frac{{\sigma}_{i}(t)}{N}\right)+1
\end{align}
Proceeding similarly all the way back to time $t=1$ gives
\begin{align}\label{eq:11}
 \Delta_i(t\!+\!1)\!=\!\Delta_i(1) \prod_{j=1}^t \left(1\!-\!\frac{{\sigma}_{i}(j)}{N}\right) 
 \!+\! \sum_{\ell=2}^{t} \prod_{j=\ell}^t \left(1\!-\!\frac{{\sigma}_{i}(j)}{N}\right)\!+\!1
\end{align}

\begin{figure}
	\centerline{\includegraphics[width=0.8\columnwidth]{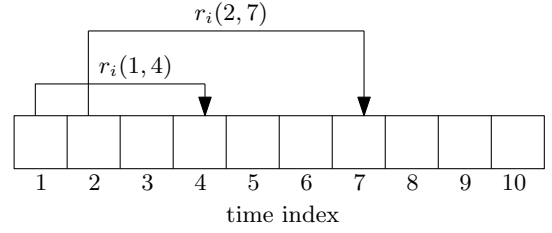}}
	\caption{Example trains for user $i$, $r_i(1,4)$ and $r_i(2,7)$.}
	\label{fig:eq9}
	\vspace*{-0.4cm}
\end{figure}

Next, we define a \emph{train} $r_{i}(k,\ell)$ for user $i$ as a series of time slots starting at slot $k$ and going up to slot $\ell$. There are $(\ell-k+1)$ elements in the train $r_i(k,\ell)$, and each train can be uniquely identified with its starting time and end time. As an example, Fig.~\ref{fig:eq9} shows trains $r_i(1,4)$ and $r_i(2,7)$ with arrows, where the starting time of the train is the starting point of the arrow and the end time is the end point of the arrow. Next, we define the \emph{value of a train} $\Gamma_i(k,\ell)$ as the multiplication of the terms $\left(1-\frac{{\sigma}_{i}(j)}{N}\right)$ for time slots $j$ in the train $r_i(k,\ell)$,  
\begin{align} \label{def:train}
  \Gamma_i(k,\ell)=\prod_{j=k}^\ell \left(1-\frac{{\sigma}_{i}(j)}{N}\right)
\end{align}
Note that each multiplicative term in the train is either $1$ if $\sigma_{i}(j)=0$ (i.e., adversary does not jam slot $j$ for user $i$) or $\frac{N-1}{N}$ if $\sigma_i(j)=1$ (i.e., adversary jams slot $j$ for user $i$). 

We can write $\Delta_i(t+1)$ in (\ref{eq:11}) using $\Gamma_i(k,\ell)$ in (\ref{def:train}) as,
\begin{align}\label{eq:12}
 \Delta_i(t+1)=\Delta_i(1) \Gamma_i(1,t)
 + \sum_{\ell=2}^{t} \Gamma_i(\ell,t)+1
\end{align}
Further, noting that $\Delta_i(1)=1$, (\ref{eq:12}) becomes
\begin{align}\label{eq:13}
 \Delta_i(t+1)= \sum_{\ell=1}^{t} \Gamma_i(\ell,t)+1
\end{align}
That is, the average age at time $(t+1)$ is the sum of the values of $t$ trains each starting from $\ell$ where $\ell=1,\ldots,t$ and ending at $t$, and further, the value of each train is a multiplication of a sequence of values $\frac{N-1}{N}$ or $1$ at each time instance depending on whether the adversary jams or not jams, respectively. 

As noted above, from (\ref{eq:13}), the average age in time slot $(t+1)$, $\Delta_i(t+1)$, depends on the adversarial actions from time slot $1$ to time slot $t$. To simplify our upcoming proofs, we assume that (\ref{eq:13}) is the expression for the average age in time slot $t$, i.e., $\Delta_i(t)$, and not time slot $(t+1)$, i.e., $\Delta_i(t+1)$. That is, we shift the age of any time slot to its previous time slot, without loss of optimality. This new assumption does not change the original problem (\ref{eq:prob_formu}), as the age at time $1$ is $1$, which is independent of any adversarial actions.

If a sequence $\sigma_{i}(t)$, $t\in\{1,\ldots,T\}$, has consecutive zeros in an interval, then we call that ${\sigma}_{i}$ a \emph{consecutive blocking sequence} (CBS). The number of ones to the left of the zeros of a CBS ${\sigma}_{i}$ is denoted as $L({\sigma}_{i})$ and the number of ones to the right side of the zeros of a CBS ${\sigma}_{i}$ is denoted as $R({\sigma}_{i})$. 

In the following lemma, we prove that for two CBSs with equal number of zeros, if the number of ones to the left of the first CBS is equal to the number of ones to the right of the second CBS, then these CBSs yield the same  average age. 
\begin{lemma}\label{lemma:3}
Let $\bar{\sigma}_{i}$ and $\Tilde{\sigma}_{i}$ be two CBSs with equal number of zeros, and $L(\bar{\sigma}_{i})=R(\Tilde{\sigma}_{i})$, $L(\Tilde{\sigma}_{i})=R(\bar{\sigma}_{i})$. Then, $\bar{\Delta}_{i}  = \Tilde{\Delta}_{i}$.
\end{lemma}

\begin{Proof}
We first note that
\begin{align}\label{eq:14}
 \sum_{t=1}^T  \sum_{\ell=1}^{t} \Gamma_i(\ell,t)
 = \sum_{t=1}^T  \sum_{\ell=t}^{T} \Gamma_i(t, \ell)
\end{align}
The equality in (\ref{eq:14}) can be interpreted as follows: From (\ref{eq:13}), the average age at time slot $t$ is a sum of $t$ trains, where the starting points of these trains vary from $1$ to $t$ and the end points of the trains is fixed at $t$. Now, (\ref{eq:14}) shows that the same sum can be written as a sum of trains whose starting points are fixed at $t$ and the end points vary from $T, T-1,\ldots,t$. Exchanging the starting/end points of these latter set of trains, they may be viewed as starting at $T, T-1,\ldots,t$ and ending at $t$ moving backwards. Thus, the time can be thought of running from $1$ to $T$, and equivalently running from $T$ to $1$. Since the second CBS is equivalent to the first CBS when the time runs from $T$ to $1$, both CBSs yield the same age for the user. 
\end{Proof}
Next, in the following lemma, we prove that if we move a CBS to right or left in such a way to increase the minimum of the number of ones on the left or the right, in other words, if we \emph{center} the CBS better, then the average age increases. That is, an adversary which \emph{centers} the CBS better over the time horizon causes more harm to the system.

\begin{lemma}\label{lemma:4}
Let CBS $\bar{\sigma}_{i}$ yield average age $\bar{\Delta}_{i}$. Let us create a new CBS  $\Tilde{\sigma}_{i}$ by applying a circular shift on $\bar{\sigma}_{i}$ to the right (or left) by one time slot. Let this new CBS yield average age $\Tilde{\Delta}_{i}$. If $\min(L(\bar{\sigma}_{i}),R(\bar{\sigma}_{i}))\leq\min(L(\Tilde{\sigma}_{i}),R(\Tilde{\sigma}_{i}))$, then $\bar{\Delta}_{i} \leq \Tilde{\Delta}_{i}$.
\end{lemma}

\begin{figure}
  \centerline{\includegraphics[width=0.7\columnwidth]{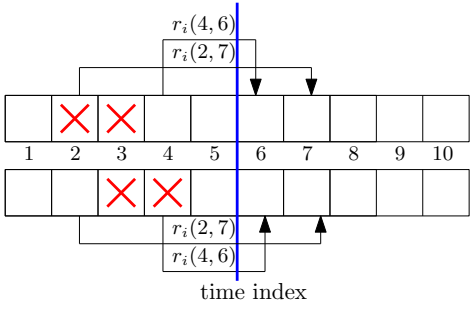}}
  \caption{Pictorial representation of the proof of Lemma~\ref{lemma:4}. Top sequence denotes CBS $\bar{\sigma}_i$ and the bottom sequence denotes CBS $\tilde{\sigma}_i$ which is shifted to the right by one. Till the vertical line both CBSs generate the same average age. After the vertical line, the average age generated by the bottom CBS is larger.}
  \label{fig:4}
  \vspace*{-0.4cm}	
\end{figure}

\begin{Proof}
Without loss of generality, consider that we circularly shift $\bar{\sigma}_{i}$ to the right by one time slot and the condition $\min(L(\bar{\sigma}_{i}),R(\bar{\sigma}_{i}))\leq\min(L(\Tilde{\sigma}_{i}),R(\Tilde{\sigma}_{i}))$ holds true. This implies that $R(\bar{\sigma}_{i})\geq L(\bar{\sigma}_{i})+1$, which further implies $R(\Tilde{\sigma}_{i})\geq L(\bar{\sigma}_{i})$. Thus, we can choose a time slot $t_{1}\in\{0,1,\ldots,T\}$ such that the whole zero portions of both of the CBSs lie in the interval $[0,t_{1}]$, and the number of $1$s to the right of $\Tilde{\sigma}_{i}$ till time $t_{1}$ is the same as $L(\bar{\sigma}_{i})$. From Lemma~\ref{lemma:3}, it follows that $\sum_{t=1}^{t_{1}}\bar{\Delta}_{i}(t)  = \sum_{t=1}^{t_{1}} \Tilde{\Delta}_{i}(t)$. A pictorial representation of this case is given in Fig.~\ref{fig:4}, where till the vertical line, which represents $t_{1}$, both sequences give the same average age. Now, for each time slot $t_{1}\leq t \leq T$, the train $r_i(j,t)$, $t_{1}\leq j \leq t$, corresponding to $\Tilde{\sigma}_{i}$ has equal or larger number of $0$s compared to the train with the same starting and end times corresponding to ${\bar{\sigma}_{i}}$. Note that, for two equal-length trains, the train with more $0$s (more blockage) provides more average age, as depending on $\sigma_{i}(t)$, the multiplier in $\Gamma_i(j,t)$ at time $\tau$, where $j\leq \tau\leq t$, can take value either $1$ (if $\sigma_i(\tau)=0$) or $\frac{N-1}{N}$ (if $\sigma_i(\tau)=1$). This is shown with example trains $r_i(4,6)$ and $r_i(2,7)$ in Fig.~\ref{fig:4} for CBSs $\bar{\sigma}_i$ (top curve) and  $\tilde{\sigma}_i$ (bottom curve). Note that the number of $0$s or $1$s in the train $r_i(2,7)$ did not change, thus $\Gamma_i(2,7)$ is same for both the CBSs, while the number of zeros in the train $r_i(4,6)$ increased, thus increasing $\Gamma_i(4,6)$ going from $\bar{\sigma}_i$ to $\tilde{\sigma}_i$. These observations imply that $\sum_{t=t_{1}+1}^{T}\bar{\Delta}_{i}(t) \leq \sum_{t=t_{1}+1}^{T} \Tilde{\Delta}_{i}(t)$. Hence, $\bar{\Delta}_{i} \leq \Tilde{\Delta}_{i}$.
\end{Proof}
For a general blocking sequence, adversary may have multiple blocks of consecutive zeros, that is, it may have blocks of ones in between blocks of zeros. In the following lemma, we prove that, instead of having multiple disconnected blocks of zeros, an optimum adversary (an adversary that gives the most harm) should have a single connected block of zeros. 

\begin{lemma}\label{lemma:5}
Among all the blocking sequences which has the same number of zeros, a CBS ${\sigma}_{i}$ with either $L({\sigma}_{i})=R({\sigma}_{i})$ or $|L({\sigma}_{i})-R({\sigma}_{i})|\leq 1$ provides the maximum average age. 
\end{lemma}

\begin{Proof}
Without loss of generality, consider a blocking sequence, $\hat{\sigma}_{i}$ which has two blocks of zeros separated by a block of ones, as illustrated in Fig.~\ref{fig6}. Thus, $\hat{\sigma}_i$ is a blocking sequence which consists of two CBSs, $\bar{\sigma}_{i}$ and $\tilde{\sigma}_{i}$, as depicted in Fig.~\ref{fig6}. Let $\bar{l}$ and $\tilde{l}$ be the lengths of the blocks of zeros of $\bar{\sigma}_{i}$ and $\tilde{\sigma}_{i}$, respectively. We call the block of zeros corresponding to $\tilde{\sigma}_{i}$ in $\hat{\sigma}_{i}$, and the block of zeros corresponding to $\bar{\sigma}_{i}$ in $\hat{\sigma}_{i}$, the right block and the left block of $\hat{\sigma}_{i}$, respectively. Without loss of generality, we assume that $T-(\bar{l}+\tilde{l})$ is an even number, i.e., a CBS ${\sigma}_{i}$ is feasible with $(\bar{l}+\tilde{l})$ length of zeros and $L({\sigma}_{i})=R({\sigma}_{i})$. Note that, if $T-(\bar{l}+\tilde{l})$ is an odd number, then the other condition holds true, namely, $|L({\sigma}_{i})-R({\sigma}_{i})|\leq 1$. To prove this lemma, we prove two possible cases. 

{\it In the first case,} we assume that $L(\tilde{\sigma}_{i})-\bar{l}$ is strictly greater than $R(\tilde{\sigma}_{i})$. Now, we create another blocking sequence, $\check{\sigma}_{i}$, by applying a left circular shift by one time slot, only to the $\tilde{\sigma}_{i}$ part of $\hat{\sigma}_{i}$, which is illustrated in Fig.~\ref{fig6}. Now we show that, $\sum_{t=1}^{T} \hat{\Delta}_{i}(t) \leq \sum_{t=1}^{T} \check{\Delta}_{i}(t)$. Note that, we can think of $\check{\sigma}_{i}$, as a blocking sequence consisting of two CBSs, namely, $\bar{\sigma}_{i}$ and the left circular shifted version of $\tilde{\sigma}_{i}$, which we call $\tilde{\tilde{\sigma}}_{i}$. We define $t_{2}$ as the time slot at which the block of zeros for $\tilde{\tilde{\sigma}}_{i}$ starts. For any $t'$, $0\leq t' < t_{2}$ $\hat{\Delta}_{i}(t') = \check{\Delta}_{i}(t')$. This is true because if we consider a train $r_{i}(j,k)$, $0\leq j \leq k$, $j\leq k <t_{2}$ corresponding to $\hat{\sigma}_{i}$, and another train with the same starting and end points corresponding to $\check{\sigma}_{i}$, they have the same value $\Gamma_i(j,k)$. From the time slot $t_{2}$ onwards the average age for the two sequences differ. Note that for all the time slots $t'$, $t_{2}\leq t' < t_{2}+\tilde{l}$, $\hat{\Delta}_{i}(t') \leq \check{\Delta}_{i}(t')$, and for all the time slots $t'$, $t_{2}+\tilde{l}\leq t' \leq T$,  $\hat{\Delta}_{i}(t') \geq \check{\Delta}_{i}(t')$. The reason for this is explained in the next paragraph. 

\begin{figure}[t]
    \centerline{\includegraphics[width = 0.7\columnwidth]{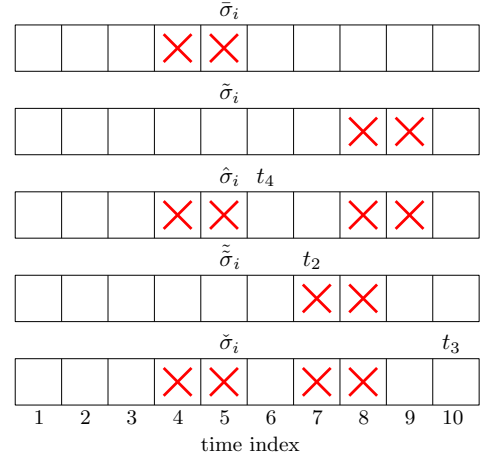}}
    \vspace*{-0.1cm}
    \caption{Pictorial representation of the proof of Lemma~\ref{lemma:5}. The original blocking sequence $\hat{\sigma}_i$ (third row) is composed of two sequences: left part $\bar{\sigma}_i$ (first row) and right part $\tilde{\sigma}_i$ (second row). The modified blocking sequence $\check{\sigma}_i$ (bottom row) is obtained by shifting the right part $\tilde{\sigma}_i$ to left by one time slot which yields $\tilde{\tilde{\sigma}}_i$ (fourth row). $\check{\sigma}_i$ is obtained by bringing two parts closer together, and yields a larger age than $\hat{\sigma}_i$. Thus, $\check{\sigma}_i$ is a better blocking sequence for the advisory than $\hat{\sigma}_i$. In this example, $\bar{l}$ and $\tilde{l}$ are equal to $2$, and  $l_{4}=2$.}
    \label{fig6}
    \vspace*{-0.5cm}
\end{figure}

Consider a time slot $t_{3}\geq t_{2}+\tilde{l}$. Note that a train corresponding to $\hat{\sigma}_{i}$ with starting point $j$, $t_{2}<j\leq t_{2}+\tilde{l}$ and ending point $t_{3}$ has always one less $\frac{N-1}{N}$ in the $(1,\frac{N-1}{N})$ pattern compared to a train corresponding to $\check{\sigma}_{i}$ with the same starting and end points. These trains are the only reasons for the fact that $\hat{\Delta}_i(t') \geq \check{\Delta}_i(t')$, where $t_{2}+\tilde{l}\leq t' \leq T$. The number of such trains for a fixed $t_{3}$ is $\tilde{l}$. For a fixed $t_{3}$, we create a set $\mathcal{T}_{\hat{\sigma}_{i},t_{3}}$ with the trains of $\hat{\sigma}_{i}$, where the starting point of those trains is $j$, $t_{2}<j\leq t_{2}+\tilde{l}$  and the end point is $t_{3}$. Let the number of ones till the time slot $t_{3}$ to the right of $\tilde{\tilde{\sigma}}_{i}$ be $l_{4}$. Due to the assumption of this case, we can always find a time slot $t_{4}$ to the left of $\tilde{\sigma}_{i}$ such that in the interval $[t_{4},t_{2}]$ we get exactly $l_{4}$ ones for $\hat{\sigma}_{i}$. Note that, a train with the starting point $t_{4}$ and the end point $j$, $t_{2}\leq j < t_{2}+\tilde{l}$ corresponding to $\check{\sigma}_{i}$ has always one less $\frac{N-1}{N}$ in the $(1,\frac{N-1}{N})$ pattern compared to a train corresponding to $\hat{\sigma}_{i}$ with the same starting and end points. The number of these trains for a fixed $t_{4}$ is again $\tilde{l}$. We create a set $\mathcal{T}_{\check{\sigma}_{i},t_{4}}$ with the trains of $\check{\sigma}_{i}$ where the starting point of those trains is $t_{4}$ and the end point of those trains is $j$, $t_{2}\leq j < t_{2}+\tilde{l}$. For each $t_{3}$ we can always find a $t_{4}$. For a fixed $t_{3}$ and $t_{4}$, for all the elements of $\mathcal{T}_{\check{\sigma}_{i},t_{3}}$, we can always find an element from $\mathcal{T}_{\hat{\sigma}_{i},t_{4}}$ which has the same number of $\frac{N-1}{N}$.  Thus, we conclude that the total average age corresponding to $\check{\sigma}_{i}$ is greater than or equal to the average age corresponding to $\hat{\sigma}_{i}$.

Now, we can move the right block of $\check{\sigma}_{i}$ further left till the assumption of this case breaks. During this process, it may happen that the right block gets connected to the left block, thus the resultant blocking sequence becomes a CBS. In this case, from Lemma~\ref{lemma:4}, we know that among all the CBSs, a CBS $\sigma_{i}$ with $R(\sigma_{i}) = L(\sigma_{i})$ achieves the maximum average age. If this does not happen (i.e., the right and left blocks do not get connected while satisfying the condition of case 1), then we continue to shift the right block of $\check{\sigma}_{i}$ till it violates the assumption of this case and we call that blocking sequence as $\hat{\hat{\sigma}}_{i}$. When it happens we are certain that the number of ones to the right of the left block of $\hat{\hat{\sigma}}_{i}$ is strictly greater than $L(\bar{\sigma}_{i})$. Now, we circularly shift the left block (i.e., $\bar{\sigma}_{i}$ part) of $\hat{\hat{\sigma}}_{i}$ to the right by one time slot. With similar arguments, we can show that the new sequence achieves higher average age. We shift the left block of $\hat{\hat{\sigma}}_{i}$ to the right till it gets connected to the right block and becomes a CBS with the same number of ones to its right and to its left. From Lemma~\ref{lemma:4}, we know that this particular CBS achieves the maximum average age.

{\it In the second case,} we consider that $L(\tilde{\sigma}_{i})-\bar{l}$ is less than $R(\tilde{\sigma}_{i})$. We circularly shift only the left block of $\hat{\sigma}_{i}$ to the right by one time slot. With similar arguments to the first case above, we can show that this new blocking sequence achieves a higher average age. We can do this right shift of the left block till it gets connected to the right block and becomes a CBS, and from Lemma~\ref{lemma:4} we know that among all the CBSs, a CBS $\sigma_{i}$ with $R(\sigma_{i}) = L(\sigma_{i})$ achieves the maximum average age. These two cases prove the statement of the lemma.
\end{Proof}
Lemma~\ref{lemma:5} implies that for a particular user $i$ if the adversary blocks $\alpha_{1} T$ slots, where $\alpha_{1}\leq\alpha$, then the adversary chooses a CBS as the blocking sequence. Next, we will prove that, if the adversary blocks only a single user with a CBS of length $\alpha T$ slots and does not block any other user, then this will give the maximum average age for the system. We state this main result in Theorem~\ref{th:1} below. Before that, we state a technical lemma, which will be used in the proof of Theorem~\ref{th:1}.

\begin{lemma}\label{lemma:6}
Consider $0\leq\beta\leq 1$, and $a,b,c$ that are integers. Then, $\beta^{a-b} - \beta^{a} \leq   \beta^{c-b} - \beta^{c} $ if $c<a$. 
\end{lemma}

\begin{Proof} 
We note
\begin{align}
\beta^{c-b} \!-\! \beta^{c} = \beta^{c}(\beta^{-b} \!- \!1) 
\geq \beta^{a} (\beta^{-b} \!-\!1) =  \beta^{a-b} \!-\! \beta^{a}
\end{align}
proving the desired result.
\end{Proof}
\begin{theorem}\label{th:1}
For the proposed randomized algorithm, the optimal blocking sequence for the adversary is to block consecutive time slots of any particular user. 
\end{theorem}

\begin{Proof}
For simplicity, we prove this in the two-user case. Let us consider the blocking matrix $\bar{\sigma}$, where the blocking length for $\bar{\sigma}_{1}$ is $\alpha_{1} T$ and the blocking length for $\bar{\sigma}_{2}$ is $\alpha_{2} T$, with $\alpha_{1} +\alpha_{2} = \alpha$. Without loss of generality, consider that the adversary blocks the first user consecutively from time slot $t_{1}$ to time slot $t_{2}$ and again consecutively from time slot $t_{3}$ to time slot $t_{4}$. Thus, $t_{2} + t_{4} - t_{1} - t_{3} + 2 = \alpha_{1} T$. Similarly, consider that the adversary blocks the second user consecutively from time slot $t_{1}'$ to time slot $t_{2}'$ and again consecutively from time slot $t_{3}'$ to time slot $t_{4}'$. Thus, $t_{2}' + t_{4}' - t_{1}' - t_{3}' +2  = \alpha_{2} T$. Without, loss of generality, we can assume that $t_{1}'<t_{2}'<t_{1}<t_{2}<t_{3}'<t_{4}'<t_{3}<t_{4}$. This is illustrated in Fig.~\ref{fig7}. 

Now, consider another blocking matrix $\tilde{\sigma}$, where the blocking length for $\tilde{\sigma}_{1}$ is $\alpha T - (t_{4}' - t_{3}' +1)$ and the blocking length for $\tilde{\sigma}_{2}$ is $(t_{4}' - t_{3}' +1 )$. For this blocking matrix, the adversary blocks the first user consecutively from time slot $t_{1}'$ to time slot $t_{2}'$ and again consecutively from time slot $t_{1}$ to time slot $t_{2}$, and finally, consecutively from time slot $t_{3}$ to time slot $t_{4}$. And, the adversary blocks the second user consecutively from time slot $t_{3}'$ to time slot $t_{4}'$ as illustrated in Fig.~\ref{fig7}. We assume that for time $t$, $\bar{\sigma}$ yields average age $\bar{\Delta}(t)$ and $\tilde{\sigma}$ yields average age $\tilde{\Delta}(t)$. Now, we prove that $\bar{\Delta}(T) \leq \tilde{\Delta}(t)$.

Consider a time slot $\bar{t}$, where $\bar{t}<t_{1}$, the train $r_{1}(j,\bar{t})$ corresponding to $\bar{\sigma}_{1}$ has the same $(1,\frac{N-1}{N})$ pattern as the train $r_{2}(j,\bar{t})$ corresponding to $\tilde{\sigma}_{2}$, where $1\leq j\leq\bar{t}$. Similarly, the train $r_{2}(j,\bar{t})$ corresponding to $\bar{\sigma}_{2}$ has the same $(1,\frac{N-1}{N})$ pattern as the train $r_{1}(j,\bar{t})$ corresponding to $\tilde{\sigma}_{1}$. Thus, we conclude that $\bar{\Delta}(\bar{t}) = \tilde{\Delta}(\bar{t})$.

\begin{figure}[t]
    \centerline{\includegraphics[width = 0.7\columnwidth]{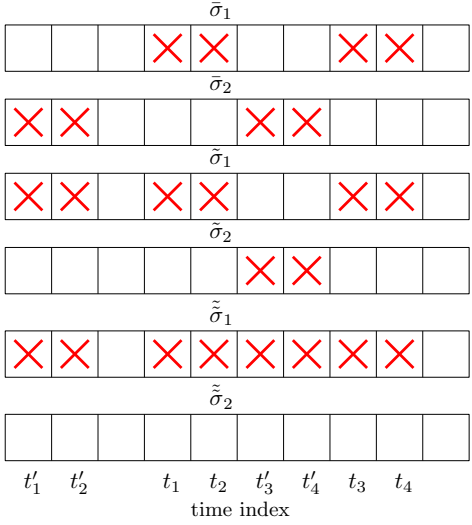}}
    \vspace*{-0.2cm}
    \caption{Pictorial representation of the proof of Theorem~\ref{th:1}. In this figure, $\bar{\sigma}_{1}$ and $\bar{\sigma}_{2}$ constitute $\bar{\sigma}$; $\tilde{\sigma}_{1}$ and $\tilde{\sigma}_{2}$ constitute $\tilde{\sigma}$; and $\tilde{\tilde{\sigma}}_{1}$ and $\tilde{\tilde{\sigma}}_{2}$ constitute $\tilde{\tilde{\sigma}}$. Till time slot $(t_{1} - 1)$, $\bar{\sigma}$ and $\tilde{\sigma}$ yield the same average age. From time slot $t_{1}$ onwards, the average age for the first user corresponding to $\tilde{\sigma}_{1}$ is higher than the average age for the first user corresponding to $\bar{\sigma}_{1}$, and the average age for the second user corresponding to $\bar{\sigma}_{2}$ is higher than the average age for the second user corresponding to $\tilde{\sigma}_{2}$. Thus, if the adversary moves from $\bar{\sigma}$ to $\tilde{\sigma}$, the average age for the first user increases and the average age for the second user decreases. Theorem~\ref{th:1} shows that this increment is more than the decrement. Thus, going from $\bar{\sigma}$ to $\tilde{\sigma}$ then to $\tilde{\tilde{\sigma}}$ increases the age.}
    \label{fig7}
    \vspace*{-0.4cm}
\end{figure}

From the time slot $t_{1}$, the average age for the two sequences differ. Consider a time slot $\bar{\bar{t}} \geq t_{1}$, the value of the train $r_{1}(j,\bar{\bar{t}})$, $\Gamma_{1}(j,\bar{\bar{t}})$ corresponding to $\tilde{\sigma}_{1}$ is $(1- \frac{1}{N})^{\bar{\bar{t}} - (x+ t_{2}' - t_{1}' + 1)} $  and similarly $\Gamma_{1}(j,\bar{\bar{t}})$ corresponding to $\bar{\sigma}_{1}$ is $ (1-\frac{1}{N})^{\bar{\bar{t}} - x} $, where $x$ is an appropriate constant depending on the time slots $\bar{\bar{t}}$ and $j$, where $1 \leq j \leq \bar{\bar{t}} $. For example, in Fig.~\ref{fig7} if  $\bar{\bar{t}}=5$ and $j=1$, then $x=2$. The value of the train $r_{2}(j,\bar{\bar{t}})$, $\Gamma_{2}(j,\bar{\bar{t}})$ corresponding to $\bar{\sigma}_{2}$ is $(1 - \frac{1} {N})^{\bar{\bar{t}} - y - (t_{2}' - t_{1}' +1)}$ and similarly, $\Gamma_{2}(j,\bar{\bar{t}})$ corresponding to $\tilde{\sigma}_{2}$ is $(1 - \frac{1} {N})^{\bar{\bar{t}} - y}$. For example in Fig.~\ref{fig7}, $j=1$ and $\bar{\bar{t}} = 5$, $y=0$. Note that for a fixed $j$ and fixed $\bar{\bar{t}}$, where $\bar{\bar{t}}\geq t_{1}$ and $j\leq\bar{\bar{t}}$, $y<x$. Thus, for a fixed $j$ and $\bar{\bar{t}}$, if the adversary chooses $\tilde{\sigma}$ over $\bar{\sigma}$ as the blocking matrix then $\Gamma_{1}(j,\bar{\bar{t}})$ increases by an amount of $ (1- \frac{1}{N})^{\bar{\bar{t}} - (x+ t_{2}' - t_{1}' + 1)} - (1-\frac{1}{N})^{\bar{\bar{t}} - x}$ and $\Gamma_{2}(j,\bar{\bar{t}})$ decreases by an amount of $(1 - \frac{1} {N})^{\bar{\bar{t}} - y - (t_{2}' - t_{1}' + 1)} -  (1 - \frac{1} {N})^{\bar{\bar{t}} - y}$. As $y<x$, from Lemma~\ref{lemma:6}, we observe that the increment of $\Gamma_{1}(j,\bar{\bar{t}})$ is more than the decrement. Thus, $\bar{\Delta}(T) \leq \tilde{\Delta}(T)$.
 
Now, we create another blocking matrix $\tilde{\tilde{\sigma}}$ where the adversary blocks the first user consecutively from time slot $t_{1}'$ to time slot $t_{2}'$, time slot $t_{1}$ to $t_{2}$, time slot $t_{3}'$ to $t_{4}'$, and finally, from time slot $t_{3}$ to $t_{4}$ and does not block the second user. With similar arguments, we can show that $\tilde{\tilde{\sigma}}$ yields a higher average age than $\tilde{\sigma}$. Finally, using Lemma~\ref{lemma:5}, we conclude that if the adversary blocks consecutive time slots for a single user, it provides the maximum average age.
\end{Proof}

Therefore, the optimum action for the adversary is to: 1) use its entire blocking power of $\alpha T$ slots, 2) block one user only, 3) block that user in consecutive $\alpha T$ slots, and 4) move the blocking sequence to the center of the time horizon $T$ as much as possible. Without loss of generality, let us assume that the adversary blocks user $1$. Now, we develop an upper bound for our proposed algorithm.

\begin{theorem}\label{th:2}
An upper bound on the average age with the proposed algorithm is $\frac{T+1}{2N} +(N-1)$.
\end{theorem}

\begin{Proof}
An upper bound for the age of the first (blocked) user is  $\frac{T(T+1)}{2}$. For other (unblocked) users, we consider that the whole horizon $T$ is divided into several intervals. We construct each interval with consecutive transmitted slots for a user followed by consecutive non-transmitted slots. Note that this is a renewal process. For any such interval, let us assume that the length of the transmitted slots is $\tau_{tr}$ and the length of the non-transmitted slots is $\tau_{ntr}$  such that $\tau=\tau_{tr}+\tau_{ntr}$. Then, 
\begin{align} 
    \sum_{\ell=1}^{\tau}\Delta_{i}(\ell) =& \sum_{\ell=1}^{\tau_{tr}}1 +\sum_{\ell=1}^{\tau_{ntr}} (1+\ell) 
    = \tau_{tr} + \frac{\tau_{ntr}^{2}}{2} + \frac{3 \tau_{ntr}}{2} \label{eq:eq10}
\end{align}
The algorithm chooses user $i$ with probability $q=\frac{1}{N}$, thus, $\tau_{tr}$ and $\tau_{ntr}$ have the following geometric distributions: 
$\mathbb{P}(\tau_{tr}=k) = q^{k-1}(1-q)$, and 
$\mathbb{P}(\tau_{ntr}=k) = q(1-q)^{k-1}$, for $k\geq 1$. 

From renewal reward theorem, we know that 
\begin{align}
\lim_{T\to \infty}\mathbb{E}\left[\frac{\sum_{t=1}^{T}\Delta_{i}(t)}{T}\right] = \frac{\mathbb{E}\left[C(\tau)\right]}{\mathbb{E}\left[\tau\right]} \label{dummy1}
\end{align}
Now from (\ref{eq:eq10}), we have
\begin{align}
    \mathbb{E}\left[C(\tau)\right] = \frac{1}{(1-q)} + \frac{3}{2q} + \frac{2-q}{q^{2}} = \frac{1}{q^{2}(1-q)}
    \label{dummy2}
\end{align}
In addition, we have
\begin{align}
    \mathbb{E}[\tau] = \mathbb{E}[{\tau_{tr}}] + \mathbb{E}[\tau_{ntr}]= \frac{1}{q(1-q)} \label{dummy3}
\end{align}
Thus, inserting (\ref{dummy2}) and (\ref{dummy3}) into (\ref{dummy1}), we obtain 
\begin{align}
\lim_{T\to \infty}\mathbb{E}\left[\frac{\sum_{i=1}^{T}\Delta_{i}(t)}{T}\right] = \frac{1}{q} = N
\end{align}
This is true for all $(N-1)$ non-blocked users. Thus,
\begin{align}
    \frac{1}{N}\sum_{i=1}^{N}\Delta_{i} \leq \frac{T+1}{2N} + (N-1)
\end{align}
completing the proof.
\end{Proof}

In the next theorem we prove the optimality of the uniform scheduling algorithm.
\begin{theorem}
The uniform scheduling algorithm is approximately $\frac{1}{\alpha^{2}}$ optimal for large enough $T$.
\end{theorem}
\begin{Proof}
From Lemma~\ref{lemma:2}, the fundamental lower bound is $\frac{T \alpha^{2}}{2 N}$, and from Theorem~\ref{th:2}, for the uniform scheduling algorithm, the upper bound is $\frac{T+1}{2N} + (N-1)$. Thus, if $\frac{T}{2 N}\gg \frac{1} {2 N} + (N-1)$, which happens when the time horizon $T$ is large enough, 
\begin{align}
\frac{\frac{T+1}{2N} + (N-1)}{\frac{T \alpha^{2}}{2 N}} \approx
\frac{\frac{T}{2N}}{\frac{T \alpha^{2}}{2 N}} \approx \frac{1}{\alpha^{2}}
\end{align}
proving the $\frac{1}{\alpha^{2}}$ optimality of the uniform user scheduling.
\end{Proof}

\subsection{Nash Equilibrium Point}
According to (\ref{eq:13}), for a scheduling algorithm $\bm{p}$ and an adversarial action $\sigma$, the expected age at time slot $(t+1)$ is 
\begin{align}\label{eq:new8}
 \Delta_i^{\bm{p},\sigma}(t+1)= \sum_{\ell=1}^{t} \Gamma_i(\ell,t)+1
\end{align}
where $\Gamma_i(\ell,t)$ is defined in (\ref{def:train}), i.e., 
\begin{align} \label{eq:new9}
  \Gamma_i(k,\ell)=\prod_{j=k}^\ell \left(1-{\sigma}_{i}(j) p_{i}\right)
\end{align}
Let us assume that the adversary blocks the middle $\alpha T$ slots consecutively for the first user; we call this adversarial action $\sigma$. Without loss of generality, we assume that $(1-\alpha)T$ is an even number, thus, $x_{1} = \frac{T} {2} - \frac{\alpha T}{2}$ is an integer. In addition, $T$ is large enough such that for a given $\alpha$, we can approximate, $(1-\alpha)T$ as a large integer. Thus, $x_{1}$ is large. From (\ref{eq:new8}) and (\ref{eq:new9}), for $1<j\leq N$,
\begin{align}
    \hspace*{-0.2cm} \Delta_{j}^{\bm{p},\sigma}= &\frac{1}{T}\Big[(T-1) (1-p_{j}) + (T-2) (1-p_{j})^{2} + \cdots + \nonumber\\
    & \quad (T -x_{1}) (1-p_{j})^{x_{1}} + \cdots + (1-p_{j})^{T-1} + T \Big] \hspace*{-0.1cm}  \\ 
    = &\frac{1}{T}\Big[T \big((1-p_{j}) + (1-p_{j})^{2} + \cdots \nonumber\\
    & \quad + (1-p_{j})^{x_{1}}+\cdots+(1-p_{j})^{T}\big)\nonumber \\
    & \quad -\big((1-p_{j}) +  2 (1-p_{j})^{2} + \cdots \nonumber\\
    & \quad + x_{1} (1-p_{j})^{x_{1}} + \cdots + T (1- p_{j})^{T}\big) + T\Big]
    \label{eq:new10}
\end{align}
For large enough $T$, (\ref{eq:new10}) can be approximated as,
\begin{align}
    \Delta_{j}^{\bm{p},\sigma} =& \frac{1}{T} \Big[T \frac{1-p_{j}}{p_{j}} - \frac{(1-p_{j})}{p_{j}^{2}} + T\Big]\\
    = & \frac{1}{T}\Big[\frac{T}{p_{j}}  - \frac{1}{p_{j}^{2}}  + \frac{1}{p_{j}}\Big] \label{eq:new11}
\end{align}
Using our assumption $T\gg\frac{1}{p_{j}}$, (\ref{eq:new11}) can be approximated as 
\begin{align}\label{eq:new12}
     \Delta_{j}^{\bm{p},\sigma} = \frac{1}{p_{j}} 
\end{align}
Similarly, from (\ref{eq:new8}) and (\ref{eq:new9}), the average age of the first user,  
\begin{align}\label{eq:new13}
    \Delta_{1}^{\bm{p},\sigma} = & \frac{1}{T}\Big[2\big((1+\alpha T) (1- p_{1})^{x_{1}} + (2+\alpha T) (1-p_{1})^{x_{1}-1} \nonumber \\
    & \quad +\cdots +(x_{1} + \alpha T) (1-p_{1})\big) + (1-p_{1})^{2} \nonumber\\
    & \quad +2 (1-p_{1})^{3} + \cdots +x_{1} (1-p_{1})^{x_{1}+1}  \nonumber \\ 
    & \quad + (x_{1}-1) (1-p_{1})^{x_{1}+2} +\cdots+ (1-p_{1})^{2 x_{1}} \nonumber\\
    & \quad + \frac{\alpha T (1+\alpha T)}{2}+T\Big]
\end{align}
Note that $x_{1}+\alpha T = \frac{T} {2} + \frac{\alpha T} {2}$. For large $T$, with similar manipulations used to get (\ref{eq:new12}), (\ref{eq:new13}) can be approximated as,
\begin{align}\label{eq:cont_eq}
    \Delta_{1}^{\bm{p},\sigma}  = (1+\alpha) \frac{1-p_{1}} {p_{1}} + \frac{\alpha (1+ \alpha T)}{2}+1
\end{align}

Thus, summing (\ref{eq:new12}) over all $j$, and (\ref{eq:new13}), the total average age for all $N$ users is
\begin{align}\label{eq:15}
    \Delta^{\bm{p},\sigma} = \frac{1}{N} \left( \sum_{j=2}^N \frac{1}{p_j} \!+\! (1\!+\!\alpha) \frac{1\!-\!p_{1}} {p_{1}} \!+\! \frac{\alpha (1\!+\! \alpha T)}{2} \!+\!1 \right)
\end{align}
The BS optimizes over $\bm{p}$ and the adversary optimizes over $\sigma$, thus, removing all the constants from (\ref{eq:15}),  $\Delta^{\bm{p},\sigma}_{eq}$, 
\begin{align}\label{eq:20}
    \Delta^{\bm{p},\sigma}_{eq} = \sum_{j=2}^{N} \frac{1}{p_{j}} +  \frac{(1+\alpha)}{p_{1}} - \alpha + \frac{\alpha (1 + \alpha T)} {2} 
\end{align}

In the following theorem, we find the probability distribution $\bm{p}$ which minimizes (\ref{eq:20}).

\begin{theorem}\label{th:new1}
For large enough $T$, if the adversary blocks user $1$ for $\alpha T$ time slots consecutively in the middle of the time horizon, then the optimal choice for the BS is to schedule the users with the following probability distribution, $p_1 = \frac{\sqrt{1+\alpha}}{N-1+\sqrt{1+\alpha}}$, and
$p_j = \frac{1}{N-1+\sqrt{1+\alpha}}$, for $j \neq 1$.
\end{theorem}

\begin{Proof}
Using the objective function in (\ref{eq:20}) together with the constraint that $\sum_{j=1}^{N} p_j=1$ yields the Lagrangian
\begin{align}
\mathcal{L} =  \sum_{j=2}^N \frac{1}{p_j} + \frac{(1+\alpha)}{p_1} +\lambda \left(\sum_{j=1}^{N} p_j-1\right)
\end{align}
The KKT conditions are
\begin{align}
\frac{1+\alpha}{p_1^2} = \lambda, \quad \frac{1}{p_j^2} = \lambda, \ j \neq 1
\end{align}
Using $\sum_{j=1}^{N} p_j=1$ to find $\lambda$ yields the optimum solution as
\begin{align}
p_1 = \frac{\sqrt{1+\alpha}}{N\!-\!1\!+\!\sqrt{1+\alpha}}, \quad 
p_j = \frac{1}{N\!-\!1\!+\!\sqrt{1+\alpha}}, \ j \neq 1
\end{align}
completing the proof.
\end{Proof}

In the following theorem, we prove the converse for Theorem~\ref{th:new1}. 

\begin{theorem}\label{th:new2}
If the BS schedules the users with a probability distribution that satisfies $p_{1}\geq p_{2}\geq \cdots\geq p_{N}$, then an optimal action for the adversary is to block user $N$ consecutively in the middle of time horizon for $\alpha T$ time slots.
\end{theorem}

\begin{Proof}
Consider that the adversary blocks the users for $\alpha_{1}T, \alpha_{2} T, \cdots,\alpha_{N} T$ time slots over the time horizon, where $\sum_{j=1}^{N} \alpha_j = \alpha$, and we denote this blocking action of the adversary as $\bar{\sigma}$. Thus, the average age of this system is ${\Delta}^{\bm{p},\bar{\sigma}}$. Note that the blocking sequences of $\bar{\sigma}$ are arbitrary, i.e., they need not be consecutive or in the middle of the time horizon. Now, we assume that the adversary blocks the users for $\alpha_{1}T , \alpha_{2}T, \cdots, \alpha_{N}T$ time slots consecutively in the middle of the time horizon, and we denote this blocking action of the adversary as $\tilde{\sigma}$. It is obvious that this adversary violates the constraint that at a time slot it can block only $1$ user. Let us denote the average age of this system as ${\Delta}^{\bm{p},\tilde{\sigma}}$. From Lemma~\ref{lemma:5}, we know that ${\Delta}^{\bm{p},\tilde{\sigma}} > {\Delta}^{\bm{p},\bar{\sigma}}$. For large $T$, similar to (\ref{eq:cont_eq}), we can approximate ${\Delta}_{i}^{\bm{p},\tilde{\sigma}}$ as $\tilde{\Delta}_{i}^{\bm{p},\tilde{\sigma}} = (1 + \alpha_{i}) \frac{1- p_{i}}{p_{i}} + \frac{\alpha_{i} (1+\alpha_{i} T)}{2}+1$. Noting that $\sum_{i=1}^N \alpha_i$ is constant and $\bm{p}$ is given and constant, we have,
\begin{align}
   \!\! \tilde{\Delta}^{\bm{p},\tilde{\sigma}}_{eq}
    = & \sum_{i=1}^{N} \frac{1}{p_{i}}  - \sum_{i=1}^{N} \alpha_{i}  + \sum_{i=1} ^{N} \frac{\alpha_{i} } {p_{i}} + \sum_{i=1}^{N} \frac{\alpha_{i} }{2} + \sum_{i=1}^{N} \frac{\alpha_{i}^{2} T}{2} \\
    = & C + \sum_{i=1}^{N} \frac{\alpha_{i}}{p_{i}} + \sum_{i=1}^{N} \frac{\alpha_{i}^{2} T } {2}
\end{align}
where $C$ is the appropriate constant. Now, if we pick any one of the $\alpha_{i}=\alpha$ and the rest of the $\alpha_{i}=0$, then this selection maximizes $\sum_{i=1}^{N} \frac{\alpha_{i}^{2}}{2}$. Let us define $k= \argmax_{i}{\frac{1}{p_{i}}}$. If the adversary chooses $\alpha_{k} = \alpha$ and the rest of the $\alpha_{i} = 0$, then this selection maximizes $\sum_{i=1}^{N} \frac{\alpha_{i}}{p_{i}}$. Let us denote this particular choice of the adversarial action as $\hat{\sigma}$. Note that $\hat{\sigma}$ is a feasible adversarial action. Let the average age of the system corresponding to $\hat{\sigma}$ be ${\Delta}^{\bm{p},\hat{\sigma}}$. Thus, ${\Delta}^{\bm{p},\hat{\sigma}}\geq{\Delta}^{\bm{p},\tilde{\sigma}}\geq{\Delta}^{\bm{p},\bar{\sigma}}$, which concludes the proof.
\end{Proof}

In the next theorem, we show that for this setting a Nash equilibrium does not exist.

\begin{theorem}\label{th:new6}
If a BS schedules $1$ user out of $N$ users at every time slot following a stationary distribution, and if an adversary can block $1$ out of $N$ communication channels between the BS and the users with a constraint that it can block maximum $\alpha T$ communication channels, and if the objective of the BS is to minimize the average age, and the objective of the adversary is to maximize the average age, then a Nash equilibrium for this problem does not exist.    
\end{theorem}

\begin{Proof}
We prove this theorem by contradiction. Let us assume that $(\bm{p}',\sigma')$ is a Nash equilibrium point for this problem, where $\bm{p}'$ is a valid probability distribution and $\sigma'$ is a valid blocking matrix. Then, according to Theorem~\ref{th:new2}, the adversary needs to block only the user which has the lowest probability of getting scheduled by the BS, for $\alpha T$ time slots in the middle of the time horizon, otherwise $\sigma'\not\in{B(\bm{p}')}$. Without loss of generality, assume that $p_{1}'\geq p_{2}'\geq\cdots \geq p_{N}'$, and the adversary blocks only user $N$ for $\alpha T$ time slots in the middle of the time horizon. From Theorem \ref{th:new1}, we know that the optimal scheduling algorithm for the BS is to schedule user $N$ with probability $\frac{\sqrt{1+\alpha}}{N-1+\sqrt{1+\alpha}}$ and schedule other users with probability $\frac{1}{N-1+\sqrt{1+\alpha}}$. Thus, user $N$ gets scheduled by the BS with the highest probability, which contradicts our assumption that $p_{1}'\geq p_{2}'\geq\cdots \geq p_{N}'$. 
\end{Proof}

\subsection{Stackelberg Equilibrium Point}
Let $\bm{p}^{*}$ be an optimal solution for the following optimization problem $\argmin_{\bm{p}\in{\mathcal{F}}} \Delta^{\bm{p}, B(\bm{p})}$. Recall that, we say the pair $(\bm{p}^{*}, \sigma^{*})$ is a Stackelberg equilibrium pair, if $\sigma^{*}\in{B(\bm{p}^{*})}$. 

In the following theorem, we find a Stackelberg equilibrium point for the system model with no diversity.

\begin{theorem}\label{new:th7}
If the BS schedules the users with uniform probability distribution and if the adversary blocks any one of the $N$ communication channels consecutively in the middle of the time horizon for $\alpha T$ time slots, then the pair of these policies is a Stackelberg equilibrium point for the communication network setting with no diversity.
\end{theorem}

\begin{Proof}
From Theorem~\ref{th:new2}, we know that if the BS schedules the users with a probability distribution $\bm{p}$ where $p_{1}\geq p_{2} \geq \cdots \geq p_{N}$, then the best action for the adversary is to block the $N$th user consecutively in the middle of time horizon for $\alpha T$ time slots. For large enough $T$, from (\ref{eq:20}), the average age of the system for this pair of actions is 
\begin{align}
     \Delta^{\bm{p},\sigma}_{eq} & =   \sum_{j=1}^{N-1} \frac{1}{p_{j}} +  \frac{(1+\alpha)}{p_{N}} - \alpha + \frac{\alpha (1 + \alpha T)} {2} \\ & = C + \sum_{i=1}^{N-1} \frac{1} {p_{i}} + \frac{(1 + \alpha)}{p_{N}}
\end{align} 
where $C$ is a constant. Thus, the optimization problem is
\begin{align}
\argmin_{\bm{p}} & \quad \sum_{i=1}^{N-1} \frac{1} {p_{i}} + \frac{(1 + \alpha)}{p_{N}} \nonumber\\
\textrm{s.t.} & \quad \sum_{i=1}^{N}p_{i} =1, \quad \text{and}
\quad p_{1} \geq p_{2} \geq\cdots\geq p_{N}
\end{align}
The Lagrangian for this problem is
\begin{align}
    \!\!\!\!\mathcal{L} \!=\! \sum_{i=1}^{N-1} \! \frac{1} {p_{i}} \!+\! \frac{(1 \!+\! \alpha)}{p_{N}} \!+\! \lambda \Big(\sum_{i=1}^{N}p_{i}\!-\!1\Big) \!+\! \sum_{i=1}^{N-1} \mu_{i} (p_{i+1}\!-\!p_{i}) \!\!
\end{align}
The KKT conditions are
\begin{align}
&p_1 = \frac{1}{\sqrt{\lambda-\mu_1}}, \qquad \quad
p_N = \frac{\sqrt{1+\alpha}}{\sqrt{\lambda+\mu_{N-1}}} \\
&p_i = \frac{1}{\sqrt{\lambda-\mu_{i-1}-\mu_i}}, \quad i=2,\ldots,N-1
\end{align}
Now, if $\mu_{N-1}=0$, then $p_{N}= \frac{\sqrt{1+\alpha}}{\sqrt{\lambda}}$ and $p_{N-1}= \frac{1}{\sqrt{\lambda + \mu_{N-2}}}$. Because of $\mu_{N-2}\geq 0$ and $\alpha> 0$, this implies $p_{N-1}< p_{N}$, which violates the primal feasibility. Thus, $\mu_{N-1}>0$. With similar arguments, $\mu_{i}>0$, for all $i=1,\cdots,N$. Now, from complementary slackness, we have $p_{1}=p_{2}=\cdots=p_{N}$.  
\end{Proof}

\section{Analysis of Age in Model With Diversity}
We consider the following scheduling algorithm. At any given time slot, the BS randomly chooses one of the $N$ users with probability  $\frac{1}{N}$, and then randomly chooses one of the $N_{sub}$ sub-carriers with probability $\frac{1}{N_{sub}}$. The BS serves the randomly chosen user in the randomly chosen sub-carrier.

\begin{theorem}\label{th:3}
For the above mentioned scheduling algorithm, the optimal blocking sequence for the adversary is to block $\alpha T$ consecutive time slots of any particular sub-carrier.
\end{theorem}

\begin{Proof}
We denote $\sum_{k=1}^{N_{sub}} \sigma_{k}(t)$ as $\tilde{\sigma}(t)$. If the adversary does not block any sub-carrier at time-slot $t$, then the age of the $i$th user at time slot $t+1$ will be $(a_{i}(t)+1)$ with probability $\frac{N-1}{N}$ and $1$ with probability $\frac{1}{N}$. If the adversary blocks any one of the $N_{sub}$ sub-carriers at time slot $t$, then the age of the $i$th user at time slot $t+1$ will be $a_{i}(t)+1$ with probability $\frac{N-1}{N} + \frac{1}{N N_{sub}}$ and $1$ with probability $\frac{1}{N} \frac{N_{sub}-1}{N_{sub}}$. Note that, if the adversary blocks any one of the $N_{sub}$ sub-carriers at time slot $t$, then $\tilde{\sigma}(t) = N_{sub} - 1$, otherwise $\tilde{\sigma}(t)=N_{sub}$. Thus, similar to the previous system model, for the $i$th user, 
\begin{align}\label{eq:basic2}
 \Delta_{i}(t + 1) = & \frac{N-1}{N} \left(\Delta_{i}(t)+1\right) + \frac{1}{N}\left(1 - \frac{N_{sub}- \tilde{\sigma}(t)}{N_{sub}}\right)   \nonumber\\ 
 & +\frac{1}{N} \left(\Delta_{i}(t)+1\right) \left(\frac{N_{sub}-\tilde{\sigma}(t)}{N_{sub}}\right)  
\end{align}
Rearranging the terms of (\ref{eq:basic2}), we get
\begin{align}\label{eq:28}
    \Delta_{i}(t+1) = \Delta_i(t) \left(1 - \frac{\tilde{\sigma}(t)}{N N_{sub}}\right) + 1
\end{align}
Note that this is multi sub-carrier version of recursion in (\ref{eq:9}).

As $\tilde{\sigma}(t)$ can take two values, either $N_{sub}$ or $N_{sub}-1$, at time slot $t$, it does not matter which sub-carrier the adversary blocks, it can block any one of the $N_{sub}$ sub-carriers, and still achieve the same $\Delta_{i}(t+1)$. Thus, we assume that the adversary always blocks one particular sub-carrier. Finally, we have to show that the adversary blocks that particular sub-carrier consecutively for $\alpha T$ time slots. Writing (\ref{eq:28}) recursively, 
\begin{align}\label{eq:29}
 \Delta_i(t+1)=&\Delta_i(1) \prod_{j=1}^t \left(1-\frac{{\tilde{\sigma}}(j)}{N N_{sub}}\right) 
 \nonumber\\
 &+ \sum_{\ell=2}^{t} \prod_{j=\ell}^t \left(1-\frac{{\tilde{\sigma}}(j)}{N N_{sub}}\right)+1
\end{align}
which is the multi sub-carrier counterpart of (\ref{eq:11}). Noting that $\Delta_{i}(1)=1$ for all $i=1,\ldots,N$, we obtain 
\begin{align}\label{eq:30}
    \sum_{i=1}^{N}\Delta_i(t+1)= N\left(\sum_{\ell=1}^{t} \prod_{j=\ell}^t \left(1-\frac{{\tilde{\sigma}}(j)}{N N_{sub}}\right)+1\right)
\end{align}

Similar to the no-diversity section, we introduce the concept of trains for ($\ref{eq:30}$). The only difference between the trains of (\ref{eq:11}) and the trains of (\ref{eq:30}) is that the elements of the trains of (\ref{eq:11}) are either $1$ or $\frac{N-1}{N}$, whereas the elements of the trains of (\ref{eq:30}) are either $\frac{N-1}{N}$ or $(1- \frac{N_{sub}-1}{N N_{sub}})$. With similar arguments to Lemma~\ref{lemma:3}, Lemma~\ref{lemma:4} and Lemma~\ref{lemma:5}, we show that blocking consecutive time slots is optimal for the adversary. 
\end{Proof}

In the next theorem, we provide an upper bound for the average age with the proposed algorithm.

\begin{theorem}\label{th:4}
The average age of the above mentioned scheduling algorithm is upper bounded by $\frac{N N_{sub}}{N_{sub}-1}$, as $T\rightarrow\infty$.
\end{theorem}

\begin{Proof}
The time horizon $T$ can be divided into intervals for a user. The intervals consist of consecutive transmitted time slots and consecutive non-transmitted time slots as discussed in the proof of Theorem~\ref{th:2}. From Theorem~\ref{th:3}, we know that the optimal blocking sequence for the adversary is to block a particular sub-carrier for consecutive $\alpha T$ time slots. Without loss of generality, we assume that the adversary blocks the first sub-carrier. Similar to the proof of Theorem~\ref{th:2}, let us assume that the length of the consecutive transmitted time slots is $\tau_{tr}$ and the length of the consecutive non-transmitted time slots is $\tau_{ntr}$. Then, $\tau_{tr}$ and $\tau_{ntr}$ are geometrically distributed as: $\mathbb{P}(\tau_{tr}=k) = q^{k-1}(1-q)$, and $\mathbb{P}(\tau_{ntr}=k)= q(1-q)^{k-1}$, for $k\geq 1$, where $q=\frac{1}{N}\frac{N_{sub}-1}{N_{sub}}$. Using renewal reward theorem,
\begin{align}
\lim_{T\to \infty}\mathbb{E}\left[\frac{\sum_{t=1}^{T}\Delta_{i}(t)}{T}\right] =\frac{1}{q}= \frac{N N_{sub}}{N_{sub}-1}
\end{align}
Thus, the average age is upper bounded as
\begin{align}\label{eq:upb}
    \Delta \leq \frac{N N_{sub}}{N_{sub}-1}
\end{align}
concluding the proof.
\end{Proof}

Next we find a universal lower bound for this system model. 

\begin{theorem}\label{th:new10}
The average expected age for the communication system with diversity is lower bounded by $\frac{N+1}{2}$, as $T\rightarrow \infty$.
\end{theorem}

\begin{Proof}
Let us assume that the adversary does not block any sub-carriers in the whole time horizon $T$ and call this adversarial action $\bar{\sigma}$. Let $\bar{\pi}_{u}$ and $\bar{\pi}_{s}$ achieve the minimum average expected age corresponding to the adversarial action $\bar{\sigma}$. Let $D_{i}(T)$ represent the number of update packets successfully received by the $i$th user in the time horizon $T$. We represent the time duration between the reception of the $j$th and $(j-1)$st update packets for the $i$th user as $I_{i}(j)$. Let $R_{i}$ be the time duration for which the $i$th user does not receive any more update packets after receiving the last packet, i.e., $\sum_{j=1}^{D_{i}(T)}I_{i}(j) + R_{i} = T$. As there is no adversarial action, the BS successfully transmits update packets in all the $T$ time slots. Thus, $\sum_{i=1}^{N} D_{i}(T) = T$.  Now, for any $T>0$,
\begin{align}
\Delta^{*} \geq \Delta^{\bar{\pi}_{u},\bar{\pi}_{s}, \bar{\sigma}} 
\end{align}
As $\bar{\pi}_{u}$ minimizes the average expected age for adversarial action $\bar{\sigma}$, $R_{i}$ must be finite almost surely, thus, $\lim_{T\rightarrow\infty} \frac{R_{i}}{T} = 0$. During the interval $I_{i}(j)$, the age of the $i$th user evolves as $\{1,2,\cdots,I_{i}(j)\}$. Thus, as $T\rightarrow\infty$, 
\begin{align}
   &\limsup_{T\rightarrow\infty} \frac{1}{NT} \sum_{i=1}^{N} \sum_{t=1}^{T}  a_{i}(t) \nonumber\\
   =& \limsup_{T\rightarrow \infty} \!\frac{1}{NT}\! \bigg(\!\sum_{i=1}^{N} \!\! \sum_{j=1}^{D_{i}(T)} \frac{I_{i}(j)(I_{i}(j)\!+\!1)}{2} \!+\! \frac{R_{i}(R_{i}\!+\!1)}{2}\!\bigg) \!\\ 
   =& \limsup_{T\rightarrow \infty} \frac{1}{2NT} \sum_{i=1}^{N} \!\sum_{j=1}^{D_{i}(T)} {I_{i}^{2}(j)} + \frac{1}{2 N T}\sum_{i=1}^{N} \!\sum_{j=1}^{D_{i}(T)} {I_{i}(j)} \\
  \geq & \limsup_{T\rightarrow \infty} \frac{1}{2N T} \sum_{i=1}^{N} \frac{1}{D_{i}(T)} \bigg(\sum_{j=1}^{D_{i}(T)} I_{i}(j)\bigg)^{2} \nonumber\\ 
  & \qquad \quad + \frac{1}{2 N T} \sum_{i=1}^{N}\sum_{j=1}^{D_{i}(T)} {I_{i}(j)} \label{step-a}\\
   = &\limsup_{T\rightarrow\infty} \frac{1}{2NT}  \sum_{i=1}^{N}\frac{1}{D_{i}(T)} (T - R_{i})^{2} \nonumber \\
   & \qquad \quad + \frac{1}{2 N T} \sum_{i=1}^{N}(T-R_{i}) \label{step-b}\\
   =& \limsup_{T\rightarrow\infty} \frac{T}{2N} \sum_{i=1}^{N} \frac{1}{D_{i}(T)} + \frac{1}{2} \label{step-c}\\ 
   =& \limsup_{T\rightarrow\infty} \frac{\sum_{i=1}^{N} D_{i}(T)}{2 N}  \sum_{i=1}^{N} \frac{1}{D_{i}(T)} + \frac{1}{2} \\ 
   \geq & \frac{N}{2}+\frac{1}{2} \label{eq:new64}
\end{align}
where (\ref{step-a}) follows from Jensen's inequality, (\ref{step-b}) follows from $\sum_{j=1}^{D_{i}(T)} I_{i}(j) + R_{i}=T$, (\ref{step-c}) follows from  $\lim_{T\rightarrow\infty}\frac{R_{i}}{D_{i}(T)} = 0$ and (\ref{eq:new64}) follows from the Cauchy–Schwarz inequality. Note,
\begin{align}
    \Delta^{\bar{\pi}_{u},\bar{\pi}_{s},\bar{\sigma}} = & \limsup_{T\rightarrow\infty} \frac{1}{NT}\sum_{i=1}^{N}\sum_{t=1}^{T} \mathbb{E}[a_{i}(t)] \\ 
    \geq & \liminf_{T\rightarrow\infty} \frac{1}{NT}\sum_{i=1}^{N}\sum_{t=1}^{T} \mathbb{E}[a_{i}(t)] \\
    \geq & \mathbb{E}\bigg[\lim_{T\rightarrow  \infty} \frac{1}{NT}\sum_{i=1}^{N}\sum_{t=1}^{T} a_{i}(t) \bigg] \label{step-a-next}\\ \geq & \frac{N}{2} +\frac{1}{2} \label{step-b-next}
    \end{align}
where (\ref{step-a-next}) follows from Fatou's lemma and (\ref{step-b-next}) follows from (\ref{eq:new64}) and noting the fact that the limit in (\ref{eq:new64}) exists.
\end{Proof}

Next, we prove the optimality of uniform user scheduling policy together with uniform sub-carrier choosing policy.

\begin{theorem}\label{th:new11}
 The uniform user scheduling algorithm together with the uniform sub-carrier choosing algorithm is $\frac{2 N_{sub}}{N_{sub}-1}$ optimal when $T\rightarrow\infty$.
\end{theorem}

\begin{Proof}
From Theorem~\ref{th:4} and Theorem~\ref{th:new10}, we have,
\begin{align}
    \frac{\frac{N N_{sub}}{N_{sub}-1}}{\frac{N}{2} + \frac{1}{2}} \leq \frac{2 N_{sub}} {N_{sub}-1}
\end{align}
which proves the $\frac{2 N_{sub}}{N_{sub}-1}$ optimality.
\end{Proof}

We note that $\frac{2 N_{sub}}{N_{sub}-1}$ is monotonically increasing in $N_{sub}$. Thus, in the worst case scenario, i.e., when $N_{sub}=2$, the uniform user scheduling together with uniform sub-carrier choosing is $4$ optimal, and for large $N_{sub}$, it is $2$ optimal.

\subsection{Nash Equilibrium Point}
In the following theorem, we find the optimal strategy for the BS for a specific blocking strategy for the adversary.

\begin{theorem}\label{th:5}
If the adversary blocks the sub-carriers uniformly and consecutively for $\alpha T$ time slots in the middle of the time horizon, then the optimal strategy for the BS is to schedule the users with uniform distribution and choose the sub-carriers with uniform distribution.
\end{theorem}

\begin{Proof}
Let the BS schedule a user among $N$ users with a pmf $\bm{p}$ and choose a sub-carrier among $N_{sub}$ sub-carriers with a pmf $\bm{q}$. Consider that the adversary is blocking a sub-carrier among $N_{sub}$ sub-carriers only for $\alpha T$ time slots with uniform distribution. Let this action of the adversary be $\sigma'$. The age in time slot $t+1$ where $\frac{T}{2} - \frac{\alpha T}{2}< t \leq\frac{T}{2} + \frac{\alpha T}{2}$ is
\begin{align}
    &\Delta_{i}^{\bm{p}, \bm{q}, \sigma'}(t+1) \nonumber\\
    = &(\Delta_{i}^{{\bm{p}, \bm{q}, \sigma'}}(t)+1)  (1-p_{i})\nonumber + p_{i}  (\Delta_{i}^{\bm{p}, \bm{q}, \sigma'}(t)+1)  \sum_{j=1}^{N_{sub}} \frac{q_{j}}{N_{sub}}  \nonumber \\ 
    & \qquad + p_{i}  \sum_{j=1}^{N_{sub}} q_{j} \left(1-\frac{1}{N_{sub}}\right)  \\ 
    = & (\Delta_{i}^{\bm{p}, \bm{q}, \sigma'}(t)+1) (1-p_{i}) + \frac{p_{i} (\Delta_{i}^{\bm{p}, \bm{q}, \sigma'}(t)+1)}{N_{sub}} \nonumber \\ 
    & \qquad + p_{i}\left(1-\frac{1}{N_{sub}}\right) \\
    = & (\Delta_{i}^{\bm{p}, \bm{q}, \sigma'}(t)+1) \left(1 - p_{i}\left(1-\frac{1}{N_{sub}}\right)\right) +p_{i}\left(1-\frac{1}{N_{sub}}\right) \\
    = &  \Delta_{i}^{\bm{p}, \bm{q}, \sigma'}(t)\left(1-p_{i}\left(1-\frac{1}{N_{sub}}\right)\right)+1
\end{align}
For $t\leq \frac{T}{2} -\frac{\alpha T}{2}$ and for $t> \frac{T}{2} + \frac{\alpha T}{2}$, the average age for time slot $t+1$ is 
\begin{align}
    \Delta_{i}^{\bm{p}, \bm{q}, \sigma'}(t+1) = \Delta_{i}^{\bm{p}, \bm{q}, \sigma'}(t) (1-p_{i}) + 1
\end{align}

In the communication network with no diversity, the adversary completely blocked one user in the middle of the time horizon for $\alpha T$ time slots. In the current model, the probability of getting scheduled by the BS for user $i$ is reduced to $p_{i}\left(1-\frac{1}{N_{sub}}\right)$ from $p_{i}$ for $\alpha T$ time slots in the middle of the time horizon $T$. Using (\ref{eq:8}) and (\ref{eq:9}) and using similar approximations to those in obtaining (\ref{eq:cont_eq}) from (\ref{eq:13}), we have,
\begin{align}\label{eq:31}
    \!\!\!\!\Delta_{i}^{\bm{p}, \bm{q}, \sigma'} =&  \frac{2 x_{1} (1-p_{i})}{p_{i}} + \frac{\alpha T \left(1- p_{i} \left(1-\frac{1}{N_{sub}}\right)\right) } {p_{i} \left(1-\frac{1}{N_{sub}}\right)} + T \\
    =&  \frac{2 x_{1}}{p_{i}} + \frac{\alpha T} {p_{i} \left(1-\frac{1}{N_{sub}}\right)}
\end{align}
where $x_{1} = \frac{T}{2} - \frac{\alpha T}{2}$. Note that, for each user, this system can be thought of as three blocks with the absence of the adversary in each block. The first block is from time slot $1$ to time slot $\frac{T}{2} - \frac{\alpha T}{2}$ with probability of getting scheduled by the BS $p_{i}$. The second block is from time slot $\frac{T}{2} - \frac{\alpha T}{2} + 1$ to time slot $\frac{T}{2} + \frac{\alpha T}{2}$ with probability of getting scheduled by the BS $p_{i}\left(1-\frac{1}{N_{sub}}\right)$. The third block is from time slot $\frac{T}{2} + \frac{\alpha T}{2} +1$ to time slot $T$ with probability of getting scheduled by the BS $p_{i}$. Thus, we get 
\begin{align} \label{eq:39}
    \Delta^{\bm{p}, \bm{q} , \sigma'} =& \sum_{i=1}^{N} \frac{(1-\alpha)T}{p_{i}} + \frac{\alpha T} {p_{i}\left(1 - \frac{1}{N_{sub}}\right)} = \sum_{i=1}^{N} \frac{c}{p_{i}} 
\end{align}
where $c$ is a constant. Minimizing (\ref{eq:39}) with respect to $\bm{p}$ subject to $\sum_{j=1}^N p_j=1$, we obtain $p_{i}=\frac{1}{N}$. As (\ref{eq:39}) does not depend on $\bm{q}$, we can choose any valid probability distribution, and we choose the uniform distribution.
\end{Proof}

The next theorem gives the optimal blocking sequence for the adversary when the BS schedules the users and chooses the sub-carriers uniformly.

\begin{theorem}
The triplet $(\bar{\bm{p}}, \bar{\bm{q}}, \sigma')$ is a Nash equilibrium for the model with diversity, where $\bar{\bm{p}}$, $\bar{\bm{q}}$ are uniform distributions over $N$ users and $N_{sub}$ sub-carriers, respectively, and $\sigma'$ blocks the sub-carriers uniformly and consecutively for $\alpha T$ time slots in the middle of the time horizon.
\end{theorem}

\begin{Proof}
From Theorem~\ref{th:3}, when the BS employs uniform user scheduling and uniform sub-carrier choosing algorithms, the optimal adversarial action is to block $\alpha T$ time slots in the middle of time horizon $T$. The adversary can choose any sub-carrier/sub-carriers in those $\alpha T$ time slots and it results the same average expected age. Thus, we assume that the adversary chooses the sub-carriers in those $\alpha T$ time slots uniformly. Thus, $\sigma' \in B(\bar{\bm{p}},\bar{\bm{q}})$. From Theorem~\ref{th:5},  $(\bar{\bm{p}},\bar{\bm{q}})\in B(\sigma')$. Thus, $(\bar{\bm{p}}, \bar{\bm{q}}, \sigma')$ is a Nash equilibrium point. 
\end{Proof}

\section{General setting for Model Without Diversity}\label{sec:5}
In this section, we extend the system models considered in the previous sections, by allowing the BS schedule multiple users and the adversary jam multiple users, at a time, respectively. In particular, for the system without diversity, at each time slot, the adversary can block communication channels of $k_{a}$ users, and the BS can schedule $k$ users. There are $n\choose k$ ways in which the BS can choose a group of $k$ users. Let the probability with which the BS chooses these $n\choose k$ groups be $p'_{1}, p'_{2}, \cdots, p'_{n\choose k}$. Note that $\sum_{i=1}^{n\choose k} p_{i}' = 1$. 

\begin{lemma}\label{lemma:ext2}
Let the probability with which the BS chooses user $i$ be $p_i$. Then, $\sum_{i=1}^{N} p_{i} = k$.
\end{lemma}

\begin{Proof}
Note that user $i$ is in $n-1 \choose k-1$ groups. Let us assume that the set $\mathcal{L}_{i}$ is the set of the indices of the groups to which the $i$th user belongs. Thus, $p_{i} = \sum_{k\in \mathcal{L}_{i}} p'_{k}$. Now, note that if we sum all the $p_{i}$s, each group comes  exactly $k$ times in that summation as each group has $k$ users. Thus, $\sum_{i=1}^{N} p_{i} = k$.   
\end{Proof}

Let us consider that the adversary chooses to block only one user (without loss of generality, assume that it is the first user) for $\alpha T$ time slots in the middle of the time horizon. In the next theorem, we determine the optimal choice for the BS for this particular blocking sequence.

\begin{theorem}\label{ref:theorem 4}
For large enough $T$, if the adversary blocks user $1$ for $\alpha T$ time slots consecutively in the middle of the time horizon, then the optimal choice for the BS is to schedule the users with the following probabilities, 
\begin{align} \label{thm14-eqn}
p_{1} = \frac{k} {1+ \sqrt{\frac{N-1}{1+\alpha}}}, \qquad p_{i} = \frac{k-p_{1}}{N-1}, \quad i\neq1 
\end{align}
\end{theorem}

\begin{Proof}
From (\ref{eq:15}), we know that for large enough $T$, the average age for the first user is $\Big((1+\alpha) \frac{1- p_{1}}{p_{1}} + \frac{\alpha  (1 + \alpha T )}{2} + 1\Big)$ and the average age of  the $i$th user, where $i\neq 1$, is $\frac{1}{p_{i}}$. Thus,
\begin{align}
    \!\!\Delta = \frac{1}{N} \bigg( \sum_{j=2}^N \frac{1}{p_j} + (1+\alpha) \frac{1-p_{1}} {p_{1}} + \frac{\alpha (1+ \alpha T)}{2} +1 \bigg)
\end{align}
Following steps similar to those in the proof of Theorem~\ref{th:new1}, we have the probabilities in (\ref{thm14-eqn}).
\end{Proof}

Theorem~\ref{ref:theorem 4} is an extended version of Theorem~\ref{th:new1}. Note that in (\ref{thm14-eqn}), $p_{1}> p_{i}$, for all $i\neq 1$. In the next theorem, we determine the optimal adversarial action corresponding to a stationary randomized scheduling policy. 

\begin{theorem}\label{thm15}
If the BS schedules the users with probabilities such that $p_{1}\geq p_{2}\geq \cdots\geq p_{N}$, then the optimal choice for the adversary is to block user $N$ in the middle of time horizon consecutively for $\alpha T$ time slots.
\end{theorem}

The proof of Theorem~\ref{thm15} follows the proof of Theorem~\ref{th:new2}. 

Next, we discuss the Nash equilibrium for this generalized setting without diversity. 

\begin{theorem}\label{thm16}
If a BS can schedule $k$ users simultaneously out of $N$ users following a stationary distribution, if an adversary can block simultaneously $k_{a}$ out of $N$ communication channels between the BS and the users with a constraint that it can block at most $\alpha T$ communication channels, and if the objective of the BS is to minimize the average age and the objective of the adversary is to maximize the average age, then a Nash equilibrium for this problem does not exist.    
\end{theorem}

The proof of Theorem~\ref{thm16} follows from the arguments made in the proof of Theorem~\ref{th:new6}. 

Next, we discuss a possible Stackelberg equilibrium point for this generalized setting when the BS acts as a leader. 

\begin{theorem}\label{thm17}
If the BS schedules the users uniformly, i.e., user $i$ gets scheduled with probability $\frac{k}{N}$ and if the adversary blocks any one of the $N$ communication channels consecutively in the middle of the time horizon for $\alpha T$ time slots, then the pair of these policies is a Stackelberg equilibrium point.
\end{theorem}

The proof of Theorem~\ref{thm17} can be constructed by making similar arguments to those in the proof of Theorem~\ref{new:th7}.

\section{Conclusion}
In this paper, we studied the interactions between an adversary and a BS in a status update system consisting of $N$ users, where the BS aims to minimize the age of the system by delivering fresh update packets to the users, and the adversary aims to increase the age of the system by hindering the transmission of these update packets via jamming. We considered two different network settings, namely, a system without any diversity and a system with diversity in the form of multiply frequency bands. For the system without diversity as well as for the system with diversity, we first found an optimal adversarial action corresponding to any randomized policy employed by the BS. Then, we proceeded to show that there does not exist any Nash equilibrium for the system without diversity, however, we find a Stackelberg equilibrium point when the BS acts as the leader. Next, we showed that a Nash equilibrium exists for the system with diversity, and we found the corresponding Nash equilibrium. Finally, we concluded the paper by  discussing results for a general setting for the system without diversity where the BS can serve multiple users and the adversary can jam multiple users, at once. 

\bibliographystyle{unsrt} 
\bibliography{references}

\end{document}